\documentclass[%
 aip,
rsi,%
 amsmath,amssymb,
 reprint,%
]{revtex4-1}

\usepackage{graphicx}
\usepackage{dcolumn}
\usepackage{bm}

\usepackage{epstopdf, epsfig}
\usepackage{subcaption}
\usepackage[section]{placeins}
\usepackage{flafter}

\begin{document}

\preprint{AIP/123-QED}

\title{A new kinetic theory model of granular flows that incorporates particle stiffness}

\author{Yifei Duan}
 \altaffiliation{Now at the Department of Chemical and Biological Engineering, Northwestern University, Illinois 60208, USA.}%
\author{Zhi-Gang Feng}%
\email{zhigang.feng@utsa.edu.}
\affiliation{ 
Department of Mechanical Engineering, UTSA, San Antonio, Texas 78249, USA
}%

\date{\today}

\begin{abstract}

Granular materials of practical interests in general have finite stiffness, therefore the particle collision is a process that takes finite time to complete. Soft-sphere Discrete Element Method (DEM) simulations suggest that there are three regimes for granular shear flows: inertial regime (or rapid flow regime), elastic regime (or quasistatic regime), and the transition regime (or elastic-inertial regime). 
If we use $t_f$ to represent the mean free flight time for a particle between two consecutive collisions and $t_c$ to represent the binary collision duration, these regimes are implicitly related to the ratio $t_c/t_f$. 
Granular flows can be successfully predicted by the classical Kinetic Theory (KT) when they are in the inertial regime of low particle-particle collision frequencies and short time contacts ($t_c/t_f \approx 0$). However, we find KT becomes less accurate in the transition regime where collision duration $t_c$ is no longer small compared with the collision interval $t_f$ ($t_c/t_f>0.05$).  
To address this issue, we develop a soft-sphere KT (SSKT) model that takes particle stiffness $k$ as an input parameter since $t_c/t_f$ is mainly determined by $k$. 
This is achieved by proposing a modified expression for the collision frequency and introducing an elastic granular temperature $T_e$. Compared with the classical KT that only considers the kinetic granular temperature $T_k$, a redefined total granular temperature ($T_g=T_k+T_e/3$) that takes both kinetic fluctuation energy and elastic potential energy into consideration is used in the SSKT model. The model is developed for identical frictionless particles with the linear-spring-dashpot (LSD) collision scheme, however, it can be extended to frictional systems as well after the modification of constitutive equations.
We show that the proposed SSKT extends the applicability of the KT framework to the transition regime without losing significant accuracy. The rheological crossover has been explained physically and the regime boundaries that separate the inertial regime and the elastic regime are quantitatively determined, showing good agreement with the previous regime map that was based on the DEM simulations. Our SSKT predictions also show that for unsteady flows such as homogeneous cooling, the particle stiffness could have a large impact on the granular flow behavior due to the energy transfer between $T_e$ and $T_k$.

\end{abstract}

\pacs{Valid PACS appear here}
\maketitle

\section{introduction}

Continuum modeling of granular and gas-solid flows generally involves the use of Kinetic Thoery (KT) models for the particle phase \citep{lun1984kinetic,jenkins1983theory}. Many KT models have been derived for dilute flows of rigid, frictionless spheres, which are the extensions of the classical kinetic theory of gases. One important difference between solid particles and gas molecules is that the kinetic energy is conserved in the elastic molecule collisions, but not conserved in the inter-particle inelastic collisions. KT requires inter-particle collisions to be instantaneous, admitting of only binary collisions, so in theory particles have to be rigid with an infinity spring constant $k$.  

 Granular particles typically have sizes that are much larger than one hundred microns. External fields such as gravity would have a much stronger effect on granular flows, which makes it difficult to experimentally investigate the flow behavior. Instead, simulations that use Lagrangian methods to track the motion of individual particles are often used to verify the KT model without the drawback of including gravity or other external effects. In general there are two Lagrangian methods, depending on if particle deformation is allowed during collisions. The first one is the event-driven hard sphere (HS) method, in which the simulation time step can vary according to the collision interval and the changes of the particle velocity occur as instantaneous and discontinuous events. Multi-body contact is not allowed in the HS method and this method may diverge at a high collision frequency, resulting in the so-called inelastic collapse \citep{mcnamara1994inelastic,luding1998handle}. The second approach is the soft-sphere Discrete Element Method (DEM) \citep{cundall1979discrete}. In the DEM, the collision is not an instantaneous event; it is resolved with a simulation time step that is much smaller than the collision duration. Both normal and tangential interactions in the DEM could be implemented by modeling the deformations of the particles; other effects such as cohesive and repulsive forces could also be added. Most importantly, the DEM allows multi-body collisions and it is more effective for granular flows in the regime where collisions take place frequently. Similar to the DEM, there is another method called Contact Dynamics (CD). It uses an implicit scheme that allows larger simulation time steps and a nonsmooth formulation to solve particle dynamics \citep{moreau1988unilateral,jean1999non}.

When it comes to a granular system of practical interest, particles always have finite stiffness. This means that the collision is a process that takes finite time to complete and the collision duration could be comparable to the collision interval at high collision frequency. Granular flows in this case can be very complex. Based on the DEM simulations of homogeneous shear flow, it was observed that granular flows could show either fluid-like behavior or solid-like behavior. In the early modeling work, generally two regimes are distinguished: inertial regime (or rapid-flow regime) and elastic regime (or quasi-static regime) \citep{campbell1990rapid}. The difference of flow behaviour between elastic and inertial regimes can be attributed to the micro-scale phenomena at the particle level. In the inertial regime, the collision duration is negligible compared to the collision interval and the hydrodynamic behaviours of granular flows are mainly dominated by the rapid binary collisions. Many researchers have reported that the instantaneous binary collision assumption is still valid even for relatively dense flows if the stiffness of contacts between particles is sufficiently large \citep{mitarai2003hard,mitarai2005bagnold,silbert2007rheology}; flows in this case can still be considered in the inertial regime. Despite the existence of velocity correlation for dense flows that could affect the KT predictions \citep{mitarai2005bagnold,mitarai2007velocity}, classical KT exhibits generally good agreement with DEM simulation results \citep{si2018development}. However, in the elastic flow regime, many adjacent particles are engaged in filament-like clusters that are called force chains; the collision duration becomes comparable to the collision interval and the sustained multi-particle contacts start to prevail. With elastic flows being dominated by enduring and multi-particle contacts, the collision duration can no longer be ignored. In the elastic regime, the stress is observed to be independent of the shear rate since the transport can occur between particles through the elastic waves travelling across the contact points at a rate which is governed by the elastic properties and not by the granular temperature \citep{campbell2002granular}. 

The transition between these regimes are observed to be smooth, which suggests that purely elastic or inertial flows are achieved only within certain limits \citep{chialvo2012bridging}. Many researchers constructed regime maps to better understand the regime transition.  The transition regime or the so-called elastic-inertial regime is identified for stresses scale using both elastic and inertial properties. \textcite{campbell2002granular} presented his regime map based on a series of DEM simulations of homogeneous shear flow including cases that are near the elastic limit. He adopted the dimensionless group $k/\rho d^3 \dot\gamma^2$ to help understand the rheology of the granular materials. Here, $\rho$ is the particle density, $d$ is the particle diameter, $k$ is the particle stiffness and $\dot \gamma$ is the shear rate. In general, the value of $k/\rho d^3 \dot \gamma^2$ reflects the ratio of the relative effects of the elastic contribution to the collisional contribution; the collisional effect dominates when $k/\rho d^3 \dot \gamma^2$ is large. As the shear rate increases, both elastic and inertial regimes approach the intermediate or transition regime that shows a stress scaling $\dot \gamma^a$ with $0<a<2$. As the name suggests, the flows in the transition regime show mixed characteristics of both elastic flows and inertial flows. This is different from the elastic regime where $\tau \propto \dot\gamma^0$ or inertial regime where $\tau \propto\dot\gamma^2$. Classical KT fails to predict the hydrodynamic behavior of granular flows in the transition regime. To model granular flows in the transition regime, researchers came up with an approach that is based on the dimensional analysis to identify dimensionless parameters of the problem. The solution of this approach are the algebraic relations between those parameters, and this approach is called $\mu(I)$ rheology \citep{da2005rheophysics}.

In the context of $\mu(I)$ rheology, there are two dimensionless numbers to characterize the granular flow regimes. One is the inertial number $I=\dot\gamma d/\sqrt{P/\rho}$ and the other one is the contact stiffness number $\kappa={k}/{Pd}$. $I$ indicates the ratio of inertial forces to the normal pressure $P$ and $\kappa$ compares the elastic force to the normal pressure $P$. Considering that in the framework of $\mu(I)$ rheology, $\phi$ does not change, and $\kappa/I^2=k/\rho d^3 \dot \gamma^2$.
These two dimensionless numbers $I$ and $\kappa$ are equivalent to $\phi$ and $k/\rho d^3 \dot \gamma^2$ from Campbell's work.
Both sets of dimensionless numbers can describe the granular flow regimes and they are complementary;  they are obtained from two different types of simulations: one is the pressure controlled shearing and the other is the volume controlled shearing. 
However, the contact stiffness is often assumed to be very large in $\mu(I)$ rheology by only focusing on the cases that has $\kappa>10^4$.\cite{da2005rheophysics}
At large contact stiffness number $\kappa$, the effect of particle stiffness is simplified and the inertial number $I$ becomes the only parameter that characterize the flow regimes.
A large value of  $I$ corresponds to the inertial regime while small $I$ corresponds to the elastic regime.  The $\mu(I)$ model has been proved to be valuable in terms of modeling "liquid-like" granular flows \cite{jop2006constitutive,martin2017continuum,mandal2016study}. However, without considering $\kappa$, the simplified $\mu(I)$ model cannot cover a wide range of granular flows as described in the work by \textcite{campbell2002granular} .
 It is ill-posed for both high and low inertial numbers as enduring elastic force chains and binary collisions becomes important at these limits and the dimensionless analysis fails to consider the additional physics during regime transition\citep{barker2015well}.  The rheology that can transition seamlessly between different regimes remains a challenge.

The aim of the present work is to extend the classical KT and develop a new KT model that can capture the mixed behavior of granular flows in the transition regime. Most of the existing KT models are limited to granular flows near the inertial limit.
There are primarily three factors that could affect the accuracy of these KT models as granular flows move away from the inertial limit: the reduced energy dissipation rate due to the existence of velocity correlation for dense granular shear flows, the reduced collision frequency as a result of finite particle contact duration, and the non-dissipative elastic potential energy that is caused by particle collisional deformation. Among these three factors, the first one is the consequence of particle contact inelasticity and the last two stem from the breakdown of instantaneous collision assumption due to the finite stiffness of particles. The velocity correlation factor has been well studied and it is believed to be the main factor that cause the discrepancy between the KT predictions and the DEM simulation results for dense homogeneous shear flows \citep{chialvo2013modified, vescovi2014plane}. \textcite{berzi2015steady} studied how the particle contact duration could reduce the collision frequency of soft-sphere granular systems and affect the collisional terms. They adjusted the granular temperature of homogeneous shear flows to account for velocity correlations. They have shown that for shear flows their model is able to produce results that agree well with the DEM simulations. However, the effect of particle stiffness to the transfer between the particle kinetic energy and elastic potential energy has not been considered in the KT models, though this factor is well known in the DEM simulations \citep{kondic2004elastic,shen2004internal,sun2013energy}. 

In the context of classical KT, kinetic granular temperature $T_k$, expressed as the average kinetic fluctuating energy, is defined to describe the collisional behavior due to particle movement:
\begin{equation} 
	T_k=\frac{1}{3n} \sum \limits_{i=1}^n {(\pmb c_i - \pmb u)^2}
	\label{eq1}
\end{equation}
where $n$ is the number density, $\pmb c$ is the individual particle's velocity and $\pmb u$ is the local average velocity.
Since no particle deformation is considered in the classical KT models, the granular temperature $T_g$ defined in these models is equivalent to the kinetic granular temperature $T_k$ defined in this work. The measured "granular temperature" from DEM simulations in the literature is calculated from Eq.(\ref{eq1}), which is referred as $T_k$ here. 
The lack of a second parameter to model the elastic behavior of particle deformation makes many KT models inappropriate for soft-sphere systems in the transition regime. 

In this study, we develop a soft-sphere KT (SSKT) model that relaxes the instantaneous collision assumption. We show that the SSKT can extend the applicability of the classical KT to the transition regime without losing significant accuracy.
This is achieved by the incorporation of two modifications in the SSKT model. The first one is to include the particle stiffness by introducing an elastic granular temperature $T_e$ that signifies the amount of elastic potential energy in the system. 
Similar to the definition of kinetic granular temperature, we define the elastic granular temperature as
\begin{equation}
    T_e=\frac{1}{mn}\sum_{i=1}^{n_{col}} k_n\xi_i^2
    \label{3.7}
\end{equation}
where $n_{col}$ is the average number of contacts per volume,  $n$ is the number density, $m$ is the mass of particle, $k_n$ is the normal stiffness and $\xi_i$ is the particle overlap of the $i^{th}$ collision.
An analytical solution to the elastic granular temperature is obtained that can be calculated using the ratio of particle collision time to mean flight time, $t_c/t_f$, which is explicitly related to the particle stiffness. In the inertial regime where $t_c/t_f \approx 0$, we have $T_e\approx 0$ and $T_g\approx T_k$, and the SSKT model is reduced to the classical KT model.  As $t_c/t_f$ increases, there would be a significant amount of energy exists in the form of non-dissipative elastic potential energy, and we define the granular temperature $T_g=T_k+T_e/3$. The hydrodynamic behavior of granular flow is determined by the particle velocity distribution, which is a function of $T_k$ instead of $T_g$. The SSKT model has better prediction due to its ability to differentiate $T_k$ from $T_g$ at large $t_c/t_f$.  The second modification is similar to the one proposed by \textcite{berzi2015steady}. The constitutive relations are modified to reflect the reduced collision frequency as a result of finite collision duration. A new correlation is developed for the particle collision frequency that includes the effect of solid volume fraction; this correlation can be extended to dense granular flows. The collisional terms such as pressure and shear stress are modified using this new correlation and the results are validated by the DEM simulations. We then apply the SSKT model to study the free cooling process and simple shear flows. We find the SSKT model is not only able to explain the discrepancy between the DEM simulation results and the predictions of classical KT, but also to produce results that agree well with the DEM simulations. Furthermore, instead of constructing a regime map based on the DEM simulation data, we create a new type of regime map based on the ratio of the elastic potential energy to the kinetic fluctuation energy.

\section{The soft-sphere kinetic theory model}

Application of any KT model involves the kinetic equation for the particle velocity distribution function. The single-particle velocity distribution function $f(\pmb c, \pmb r, t)$ governs the macroscopic properties of solid particles, which is a function of granular temperature. Here $\pmb c$, $\pmb r$ and $t$ are the particle velocity, position vector and time variable, respectively.
There are two granular temperatures defined in this work.
Kinetic granular temperature $T_k$ corresponds to the particle fluctuations while elastic granular temperature $T_e$ corresponds to the particle deformation.
Classical KT model is for rigid body hard-sphere systems; it has no particle deformation so $T_g=T_k$ for these systems.
The granular temperature for hard-sphere systems can refer to either $T_g$ or $T_k$ since they are equal. However, for soft-sphere systems, $T_k$ is not equal to $T_g$ and the single particle velocity distribution function $f(\pmb c, \pmb r, t)$ is determined by $T_k$ instead of $T_g$.

For dilute homogeneous shear flows under uniform shear $\dot \gamma$ without external effect, we can write the kinetic fluctuation energy balance equation as
\begin{equation}
	\frac{3}{2}\rho \phi \frac{d T_g}{dt} = \pmb \tau \dot\gamma - \Gamma
	\label{energyequation}
\end{equation}
where $\phi$ is the solid volume fraction, $\pmb \tau$ is the shear stress and $\Gamma$ is the energy dissipation rate. A complete energy balance equation for general cases and the related transport coefficients can be found in reference\citep{benyahia2012summary}.
By defining the total granular temperature $T_g=T_k+T_e/3$ on the left hand side of equation, it can capture the energy transfer between kinetic energy and elastic potential energy observed in the DEM simulations. 
If we can express $T_e$ as a function of $T_k$ , the energy balance equation can be closed and the change of $T_k$ is a result of both inelastic collision and the energy transfer from or to $T_e$.

Based on the averaging method\citep{babic1997average}, \textcite{artoni2015average} derived the balance equations to consider the conversion of mechanical energy due to affine deformations. Though they did not directly include the elastic energy in the energy balance equation, an additional energy dissipation term was introduced for the long-lasting contacts. 
\textcite{kondic2004elastic} explored the role of elasticity in a system and showed that the elastic energy could be larger than the kinetic fluctuation energy in cases of high solid volume fraction.
This portion of energy, $T_e$, cannot be dissipated during collisions but it can be transferred to $T_k$ spontaneously, which will eventually affect the bulk behavior of granular flows. 
In this section, we derive a model based on the LSD scheme in order to calculate the amount of elastic energy in a system. The LSD model is commonly used to calculate the inelastic force between colliding spheres in the DEM simulations. It leads to a constant coefficient of normal restitution. For the collision of two identical frictionless particles, the particle normal interaction during a contact is described as follows: 

\begin{equation}
  \ddot \xi=-\frac{k}{m_{eff}}\xi-\frac{\eta_n}{m_{eff}}\dot \xi
  \label{3.1}
\end{equation}
where $m_{eff}=m/2$, $m$ is the mass of the particle, $\eta_n$ is the normal damping coefficient, and $\xi$ is the normal overlap of two colliding particles. By solving Eq.(\ref{3.1}) with the initial conditions of $\xi=0$ and $\dot\xi=V$, we have
\begin{equation}
    \xi=\frac{V}{\sqrt{1-\beta^2}\omega_0}\sin (\sqrt{1-\beta^2}\omega_0 t)\exp(-\beta\omega_0t)
    \label{3.2}
\end{equation}
\begin{equation}
\begin{split}
     & \dot\xi=V\cos (\sqrt{1-\beta^2}\omega_0 t)\exp(-\beta\omega_0t)-\frac{V\beta}{\sqrt{1-\beta^2}} \\
     & \times\sin (\sqrt{1-\beta^2}\omega_0 t)\exp(-\beta\omega_0t)
\end{split}
    \label{3.3}
\end{equation}
Here, $\beta=\frac{\eta_n}{2\sqrt{k_nm_{eff}}}$ and $\omega_0=\sqrt{\frac{k_n}{m_{eff}}}$, $V$ is the pre-collision normal relative velocity at the point of contact. Given the value of the normal restitution coefficient $e$, the normal damping coefficient and binary collision duration can then be obtained, 
\begin{equation}
    \eta_n=\frac{2\sqrt{m_{eff}k_n}|\ln e|}{\sqrt{\pi^2+\ln^2e}}
    \label{3.4}
\end{equation}
\begin{equation}
  t_{c}=\pi(\frac{k_n}{m_{eff}}-\frac{\eta_n^2}{4m_{eff}^2})^{-\frac{1}{2}}
  \label{3.5}
\end{equation}
At any given time, the total elastic energy in the system is measured by
\begin{equation}
    E_e=\frac{1}{2}\sum_{i=1}^{N_{col}}k_n\xi _i^2
\end{equation}
here $N_{col}$ is the total number of collisions at this specific time, and $\xi_i$ is the overlap of $i^{th}$ collision in the system. 
The time averaged square of the normal overlap $\overline{\xi_i^2}$ during $i^{th}$ collision follows
\begin{equation}
    \overline{\xi_i^2}=\frac{1}{t_{c}}\int_0^{t_{c}} \xi_i^2dt=\frac{1}{2}(\frac{V_i}{\omega_0})^2\Big[1-\frac{\sqrt{1-\beta^2}\sin(\frac{2\pi}{\sqrt{1-\beta^2}})}{2\pi}\Big]
    \label{3.9}
\end{equation}
Here, $V_i$ is the pre-collision normal relative velocity component at the point of contact of $i^{th}$ collision. Substitute $\omega_0=\sqrt{\frac{2k_n}{m}}$ into Eq.(\ref{3.9}), we have

\begin{equation}
   {k_n\overline{\xi_i^2}}=\frac{1}{4}{mV_i^2}\Big[1-\frac{\sqrt{1-\beta^2}\sin(\frac{2\pi}{\sqrt{1-\beta^2}})}{2\pi}\Big]
   \label{3.10}
\end{equation}
For a soft-sphere system with large portion of elastic potential energy, there are a lot of collisions taking place in the system, so the sum of $\xi_i^2$ of each collision at any given time could be approximated by the sum of the time averaged value $\overline{\xi_i^2}$:
\begin{equation}
    \sum_{i=1}^{N_{col}}\xi_i^2\approx\sum_{i=1}^{N_{col}}\overline{\xi_i^2}
    \label{3.11}
\end{equation}
Then we can rewrite Eq.(\ref{3.7}) based on Eqs.(\ref{3.10}) and (\ref{3.11})
\begin{equation}
   T_e=\frac{\sum_{i=1}^{N_{col}} (k_n \overline{\xi_i^2})}{mN}=\frac{1}{4}\frac{N_{col}}{N}V_{avg}^2\Big[1-\frac{\sqrt{1-\beta^2}\sin(\frac{2\pi}{\sqrt{1-\beta^2}})}{2\pi}\Big]
   \label{3.12}
\end{equation}
where $V_{avg}^2=\frac{1}{N_{col}}\sum_{i=1}^{N_{col}}V_i^2$. According to the KT, for three dimensional systems the energy dissipation rate \citep{duan2017incorporation}
\begin{equation}
   \Gamma=\frac{12\rho\phi^2(1-e^2)g_0 T_k^{\frac{3}{2}}}{d\sqrt{\pi}}
   \label{3.13}
\end{equation}
where $\phi$ is the solid volume fraction, $e$ is the coefficient of restitution and $g_0$ is the radial distribution function \citep{torquato1995nearest}. 
We assume only binary collisions exist in the system.
The collision frequency per particle is equal to the inverse of mean free flight time between two consecutive collisions \citep{duan2017modified}, 
\begin{equation}
	t_f^{-1}=\frac{12}{d}\phi g_0 \sqrt{\frac{T_k}{\pi}}
	\label{eq.freq}	
\end{equation}
$t_f^{-1}$ can be interpreted as the average number of collisions per second for each individual particle.  The derivation of Eq.(\ref{eq.freq}) can be found in Appendix. Within one second, the total number of collisions in the system is $Nt_f^{-1}$. $\Gamma$ is defined as the energy dissipation per second, so the average energy dissipation per collision follows:
\begin{equation}
   \frac{\Gamma}{Nt_f^{-1}}=m(1-e^2)T_{k}
   \label{3.14}
\end{equation}
The average kinetic energy change during a collision could also be derived based on the restitution coefficient as follows: 
\begin{equation}
   \frac{1}{N_{col}}\sum_{i=1}^{N_{col}}\Delta E_i=\frac{1}{4}m(1-e^2)V_{avg}^2
   \label{3.15}
\end{equation}
$\Delta E_i=\frac{1}{4}m(1-e^2)V_i^2$ is the transnational kinetic energy change during a collision\citep{lun1984kinetic}. Since both expressions on the right hand side of Eq.(\ref{3.14}) and Eq.(\ref{3.15}) are the energy dissipation per collision and they must be equal, we can relate the average normal collision velocity $V_{avg}$ to the kinetic granular temperature $T_k$ by
\begin{equation}
   V_{avg}^2=4T_k
   \label{3.16}
\end{equation}
By combining Eq.(\ref{3.12}) and Eq.(\ref{3.16}), the ratio of elastic temperature to kinetic temperature can be written by
\begin{equation}
   \frac{T_e}{T_k}=\frac{N_{col}}{N}\Big[1-\frac{\sqrt{1-\beta^2}\sin(\frac{2\pi}{\sqrt{1-\beta^2}})}{2\pi}\Big]
   \label{3.17}
\end{equation}
\begin{figure}
  \centerline{\includegraphics[scale=0.4]{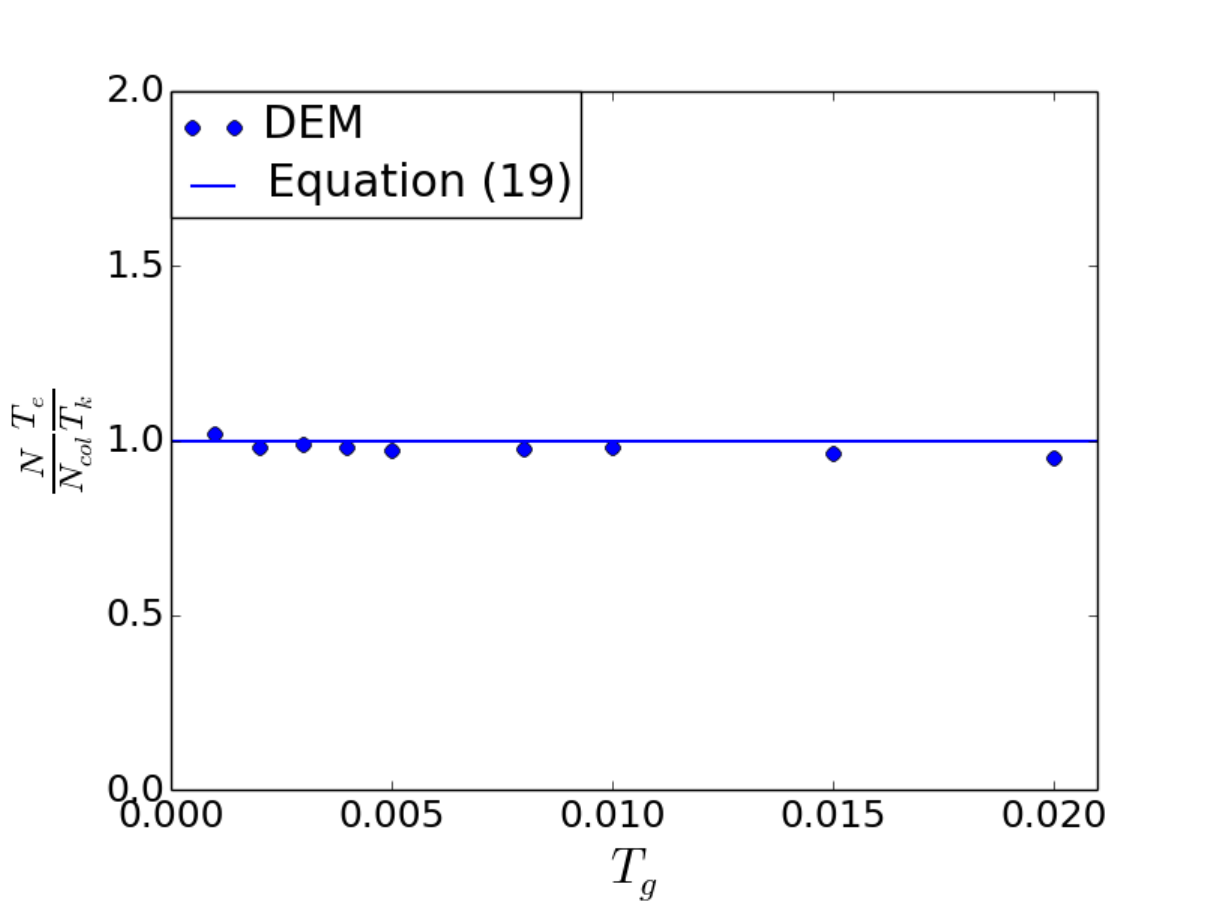}}
  \caption{$NT_e/N_{col}T_k$ at different initial granular temperature $T_g$ in $m^2/s^2$. Simulations were set up in a cubic domain using 2984 particles with periodic boundary conditions ($\phi=0.2$). Dots represents the time averaged ratio from the DEM simulations.}
\label{fig1}
\end{figure}
To verify Eq.(\ref{3.17}), we calculate the value of $NT_e/N_{col}T_k$ from the DEM simulations. The DEM simulations were performed using the MFiX package \citep{syamlal1993mfix}, which is available from the National Energy Technology Laboratory (NETL). The DEM simulation time step $\Delta t$ must be sufficiently small compared with the particle collision time $t_{c}$  in order to resolve the particle collision process. For the present simulations we chose $\Delta t=t_{c}/50$, a practice successfully employed by others \citep{silbert2001granular,chialvo2013modified, gu2016modified}.  Particles were packed into a cubic domain with periodic boundaries.
Each particle was given an initial velocity that follows the Maxwell distribution function, so the system has an initial granular temperature $T_g=T_k$. We choose $e=1$ so that without external effects, the total energy in the system remain constant and the system quickly reaches steady state. At steady state $T_e$ becomes a non-zero term and $T_g=T_k+T_e$. Both $T_e$ and $T_k$ can be calculated from DEM simulations based on their definitations.

According to Eq.(\ref{3.17}), the ratio $NT_e/N_{col}T_k$ is only related to the particle properties and it is close to 1 if the normal restitution coefficient $e\rightarrow1$. We calculated the ratio from DEM simulations of granular flows with $e=1$ at different granular temperatures and found good agreement with Eq.(\ref{3.17}), as shown in figure \ref{fig1}. Note that the derivation is not limited to $e=1$ or steady state, though rapid change of $T_e$ can slightly affect the accuracy of Eq.(\ref{3.11}).  The relation between $T_k$ and $T_e$ is still valid at $e<1$, which can be proved in the homogenous cooling case in the following section.

Since a collision is a process that takes a finite amount of time $t_{c}$ to complete, particle is considered in a collision state during this time window of $t_{c}$. If the mean free flight time of each particle is $t_f$, the probability of any particle in the collision state will be $t_c/t_f$, so we expect $N_{col}/N=t_c/t_f$ at any given time. Therefore, we can rewrite Eq.(\ref{3.17}) to
\begin{equation}
   T_e={t_{c}t_f^{-1}\Big[1-\frac{\sqrt{1-\beta^2}\sin(\frac{2\pi}{\sqrt{1-\beta^2}})}{2\pi}\Big]}T_k
   \label{3.18}
\end{equation}

\begin{figure}
\centering
\begin{subfigure}{1\columnwidth}
\includegraphics[width=\columnwidth,keepaspectratio]{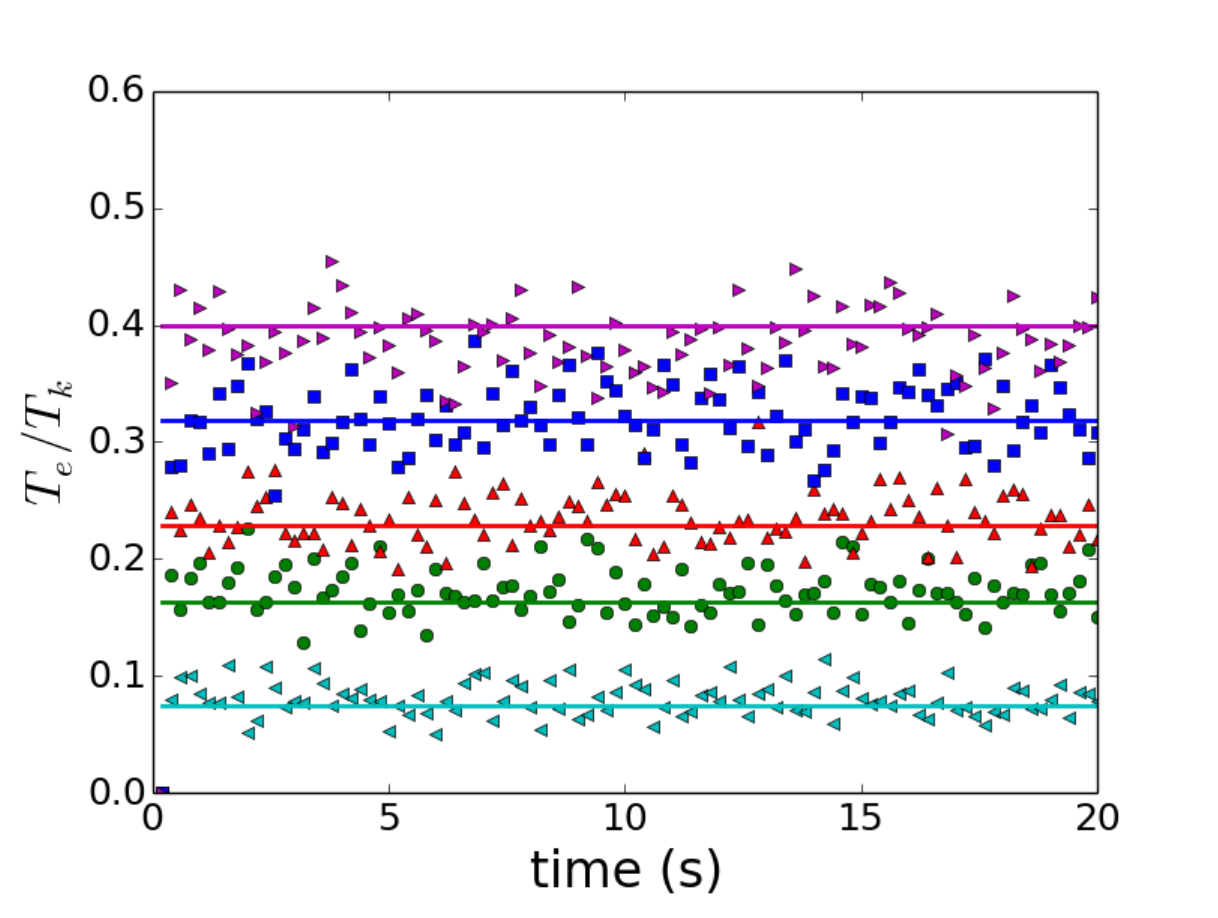}
\caption{ The ratio of the elastic temperature $T_e$ to the kinetic temperature $T_k$ in granular systems with different $t_c/t_f$. Markers represents the DEM results (from top to bottom, $t_c/t_f$=0.07, 0.16, 0.22, 0.30, 0.38) and lines are the predictions from Eq.(\ref{3.18}).}
\label{fig:4a}
\end{subfigure}\hfill
\begin{subfigure}{1\columnwidth}
\includegraphics[width=\columnwidth,keepaspectratio]{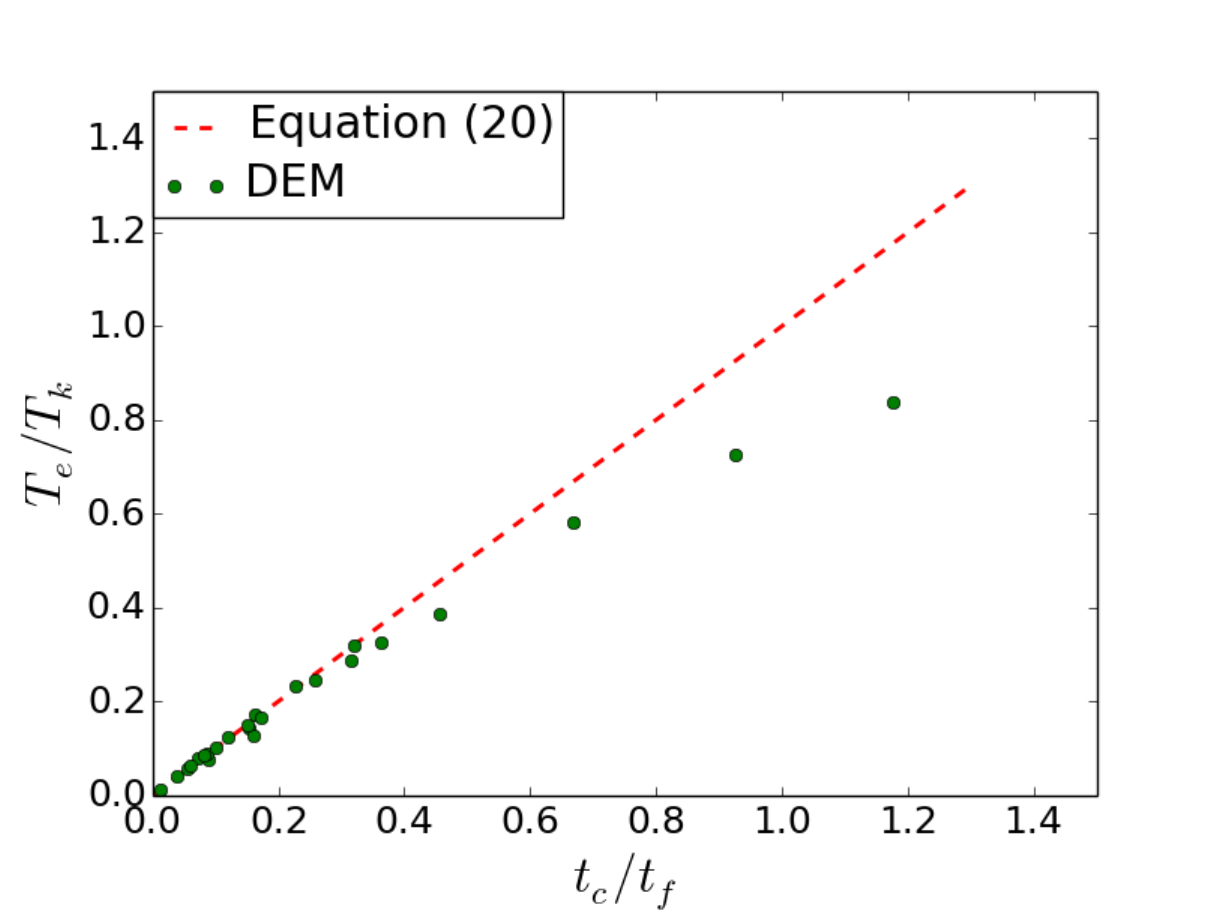}
  \caption{The value of $T_e/T_k$ along with the ratio of  binary collision time to collision interval $t_c/t_f$. Dashed line indicates the theoretical prediction; dots represent DEM results at $\phi$ ranging from 0.1 to 0.5.}

\label{fig3}
\end{subfigure}
\caption{$T_e/T_k$ compared with DEM results at different $t_c/t_f$.}
\label{fig:4}
\end{figure}

\begin{figure}
  \centerline{\includegraphics[scale=0.4]{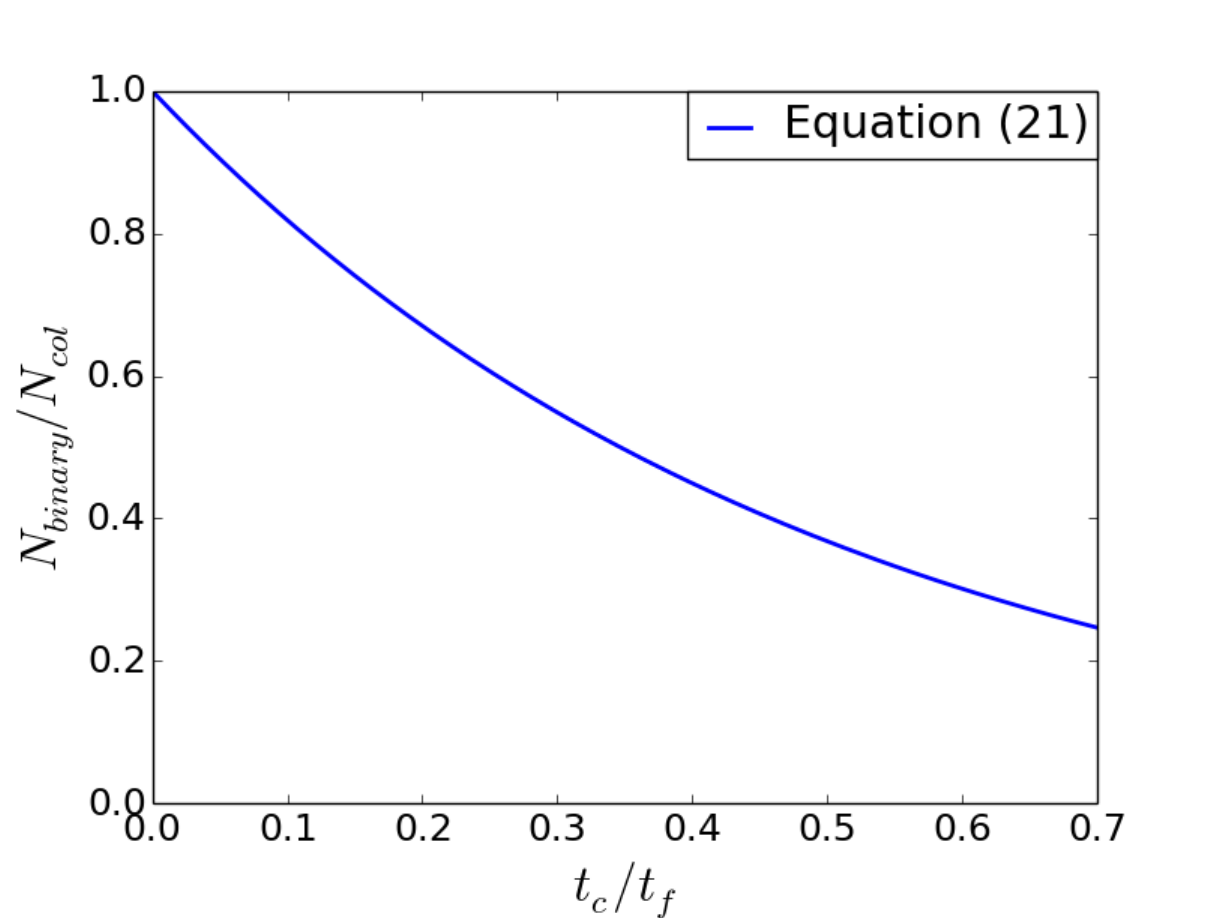}}
\caption{The ratio of  the number of binary collisions to the number of total collisions in a system. }\label{fig:4b}
\end{figure}

To verify Eq.(\ref{3.18}), DEM simulations were performed to compute the value of $T_e/T_k$ of the system. Initial particle velocities were given based on the initial granular temperature $T_g$; the values of $T_e/T_k$ were calculated as a function of time. From figure \ref{fig:4a} we can see for a system with particles of finite stiffness, our proposed model could predict the ratio of $T_e$ to $T_k$ that matches the DEM results very well. However, it starts to deteriorate as the portion of elastic energy further increases. This is because the model relies on the velocity profile that is based on the binary collision assumption. It is expected that with the increasing portion of non-binary collisions, the proposed model would become less accurate and the time averaged elastic deformation becomes more complicated. Based on the probability analysis, the binary collision percentage in a system is \citep{luding1998handle,duan2017modified} 
\begin{equation}
   P_{binary}=\exp(-2t_{c}t_f^{-1})
   \label{3.19}
\end{equation}

$T_e/T_k$ is shown to increase with $t_c/t_f$ as shown in figure \ref{fig3}. From figure \ref{fig:4b}, we find a larger $T_e/T_k$ results in a larger fraction of non-binary collisions, which reduces the accuracy of our proposed models. This is evidenced in figure \ref{fig3}. For a DEM simulation with $t_c/t_f=0.4$, the proposed model can still predict the ratio of $T_e$ to $T_k$ quite well with less than 10$\%$ difference compared to the DEM simulation results. In other word, the non-binary collisions which weight about 65$\%$ of the total collisions in the DEM simulations cause only 10$\%$ difference compared with the ideal case of all binary collisions.

\begin{figure}
\centering
\begin{subfigure}{1\columnwidth}
\includegraphics[width=\columnwidth,keepaspectratio]{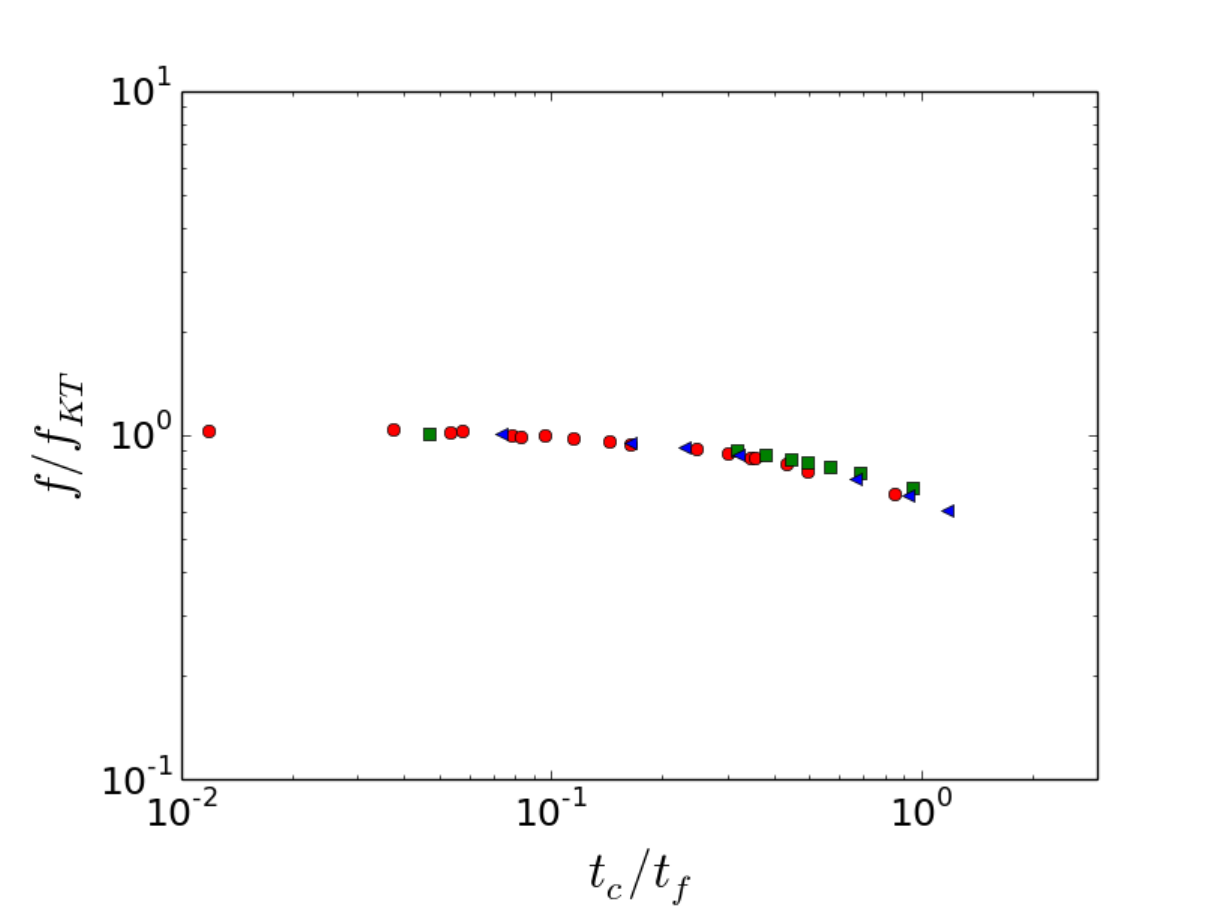}
\caption{The ratio of the measured collision frequency $f$ from DEM to the predicted collision frequency $f_{KT}$ by classical KT for granular systems at different $t_c/t_f$.}
\label{fig:freq-a}
\end{subfigure}\hfill
\begin{subfigure}{1\columnwidth}
\includegraphics[width=\columnwidth,keepaspectratio]{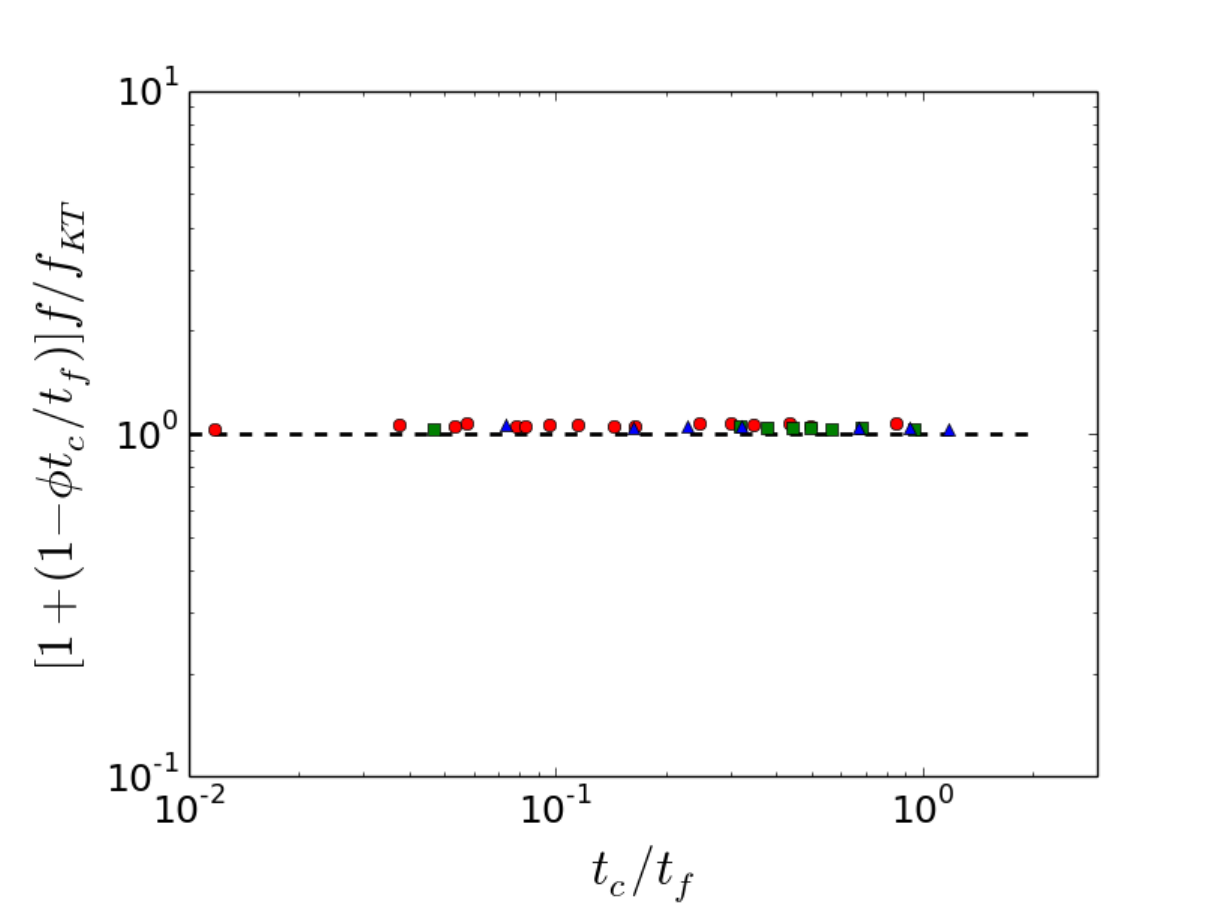}
\caption{Rescaled ratio $(1+(1-\phi) t_c/t_f)\frac{f}{f_{KT}}$ at different $t_c/t_f$. }
\label{fig:freq-b}
\end{subfigure}
\caption{Markers represent DEM results (Circle: $\phi=0.3$; Triangle: $\phi=0.4$; Square: $\phi=0.5$)}
\label{fig:freq}
\end{figure}

When it comes to the SSKT model, the single-particle velocity distribution function should be determined by $T_k$ instead of $T_g$; the existence of elastic energy should affect the fluctuation energy dissipation rate since this portion of energy could not be dissipated during collisions, but it could be transformed into the form of kinetic energy at any time. This introduces one more unknown $T_k$ and Eq.(\ref{3.18}) has to be used as a closed-form constitutive expression. 

Moreover, we found with the increase of $t_c/t_f$, the actual collision frequency is less than the value predicted by the classical KT. 
\begin{equation}
    f_{KT}=\frac{12}{d} \phi g_0 \sqrt{\frac{T_k}{\pi}}
    \label{freqsoft}
\end{equation}
Eq.(\ref{freqsoft}) represents the KT prediction of collision frequency for hard sphere systems; it yields good agreement with the hard sphere simulation results \citep{duan2017modified}.
As shown in figure \ref{fig:freq}, we find the collision frequency for soft sphere systems decreases when $t_c/t_f$ increases. Based on the DEM simulation results, we found the soft sphere collision frequency could be nicely correlated by the following equation:
\begin{equation}
    f=\frac{f_{KT}}{1+(1-\phi) t_c/t_f}
    \label{eq:f_soft}
\end{equation}
This expression resembles the one proposed by \textcite{berzi2015steady}, $f= \frac{f_{KT}}{1+ t_c/t_f}$, when the system is extremely dilute (i.e., $\phi \approx 0$). Because the collisional stress and dissipation rate of kinetic fluctuation energy are proportional to the collision frequency, the constitutive relations that are derived for hard spheres should be modified accordingly in order to be used for soft spheres. Based on the expressions of the kinetic theory for hard spheres \citep{garzo1999dense,jenkins2010dense}, the constitutive equations of SSKT model are listed in Table \ref{table1}.  The expressions for classical KT and the extended KT by \textcite{berzi2015steady} are also listed for comparison. 
\begin{table*}
   \caption{Summary of model equations.}
   \label{table1}	
   \begin{ruledtabular}
   \def\arraystretch{1.8}
   \begin{tabular}{p{5cm} p{5cm} p{5cm}}
   
   Classical KT\citep{berzi2015steady,jenkins2010dense} & \textcite{berzi2015steady} & SSKT\\
   \hline

   $ T_g=T_k$
   & $ T_g=T_k$ 
&    $ T_g=T_k+\frac{1}{3}T_e$\\

   $ T_e=0$   
   &   $ T_e=0$   
&    $ {T_e}=K H T_k^{3/2}$\\

   $ \Gamma=\frac{12}{d\sqrt{\pi}}\rho \phi^2g_0(1-e^2)T_k^{3/2}$
   & $ \Gamma=\frac{12}{d\sqrt{\pi}}\rho \phi^2g_0(1-e^2)T_k^{3/2}\frac{f}{f_{KT}}$
& $\Gamma=\frac{12}{d\sqrt{\pi}}\rho \phi^2g_0(1-e^2)T_k^{3/2}\frac{f}{f_{KT}}$\\

   $p=\rho\phi T_k+ 2\rho(1+e)\phi^2 g_0 T_k$
   &    $p=\rho\phi T_k+ 2\rho(1+e)\phi^2 g_0 T_k \frac{f}{f_{KT}}$
&    $p=\rho\phi T_k+ 2\rho(1+e)\phi^2 g_0 T_k \frac{f}{f_{KT}}$\\

   $\pmb \tau=\frac{8}{5} \rho d \phi^2  g_0 J \sqrt{\frac{T_k}{\pi}} \dot \gamma$
   & $\pmb \tau=\frac{8}{5} \rho d \phi^2  g_0 J \sqrt{\frac{T_k}{\pi}} \dot \gamma \frac{f}{f_{KT}}$
& $\pmb \tau=\frac{8}{5} \rho d \phi^2  g_0 J \sqrt{\frac{T_k}{\pi}} \dot \gamma \frac{f}{f_{KT}}$\\

   $f_{KT}=t_f^{-1}=\frac{12}{d} \phi g_0 \sqrt{\frac{T_k}{\pi}}$  
    &    $f_{KT}=t_f^{-1}=\frac{24}{d} \phi g_0 \sqrt{\frac{T_k}{\pi}}$  
 &    $f_{KT}=t_f^{-1}=\frac{12}{d} \phi g_0 \sqrt{\frac{T_k}{\pi}}$    \\
 
    $J=\frac{1+e}{2} + \frac{ \pi(1+e)^2(3e-1) }{96-24(1-e)^2-20(1-e^2)}$  
    &     $J=\frac{1+e}{2} + \frac{ \pi(1+e)^2(3e-1) }{96-24(1-e)^2-20(1-e^2)}$   
 &     $J=\frac{1+e}{2} + \frac{ \pi(1+e)^2(3e-1) }{96-24(1-e)^2-20(1-e^2)}$   \\

\\

& $t_c=\frac{d}{5}(\frac{\rho \pi d}{4k})^{\frac{1}{2}}$
& $t_{c}=\pi(\frac{k_n}{m_{eff}}-\frac{\eta_n^2}{4m_{eff}^2})^{-\frac{1}{2}}$ \\

&      $ \frac{f}{f_{KT} }= [1+\frac{t_c}{t_f}]^{-1}$
&     $ \frac{f}{f_{KT} }= [1+(1-\phi) \frac{t_c}{t_f}]^{-1}$\\


&    
&    $\eta_n=\frac{2\sqrt{m_{eff}k_n}|\ln e|}{\sqrt{\pi^2+\ln^2e}}$\\

&
&    $\beta=\frac{\eta_n}{2\sqrt{k_nm_{eff}}}$\\                

&
&     $K=\frac{12}{d }\phi g_0 \sqrt{\pi}(\frac{k}{m_{eff}}-\frac{\eta_n^2}{4m_{eff}^2})^{-\frac{1}{2}}$\\

&
&     $H=1-\frac{1}{2\pi}\sqrt{1-\beta^2}\sin(\frac{2\pi}{\sqrt{1-\beta^2}})$\\

\\

   \end{tabular}
   \end{ruledtabular}
\end{table*}

In Table.\ref{table1}, the energy dissipation rate $\Gamma$ of classical KT can be found in many KT models\cite{lun1984kinetic,jenkins2010dense}, and the pressure $p$ and shear stress $\tau$ of classical KT are obtained from equations 25 and 26 in the work by \textcite{berzi2015steady}. Note that if we use $p_{KT}$ to represent the prediction of classical KT,  the modified pressure is written as $p^{-1}=p_{KT}^{-1}+(p_{KT} \frac{tc}{tf})^{-1}$ in their extended KT model. This is equivalent to the modification that all collision frequency related terms are multiplied by a factor $f/f_{KT}$ or $1/(1+t_{c}/t_f)$, which is the form we used in Table \ref{table1}.  As an intermediate variable, collision frequency $f_{KT}$  is not explicitly derived in many KT models, and its definition may vary case by case. In our work, $f_{KT}$ is defined as the number of collisions per particle in one second. The detailed derivation can be found in the Appendix. Also the expression of $T_e$ from SSKT in Table \ref{table1} has a different form compared with Eq.(\ref{3.18}), in which $K$ and $H$ are used to replace $t_c/t_f$ ($t_c/t_f=KT_k^{1/2}$), so $T_k$ can be moved to the right hand side of equation. Due to the reduced collision frequency, terms that are derived based on particle collisions are multiplied by the $f/f_{KT}$ according to Eq.(\ref{eq:f_soft}) .

\begin{figure}
\centering
\begin{subfigure}{1\columnwidth}
\includegraphics[width=\columnwidth,keepaspectratio]{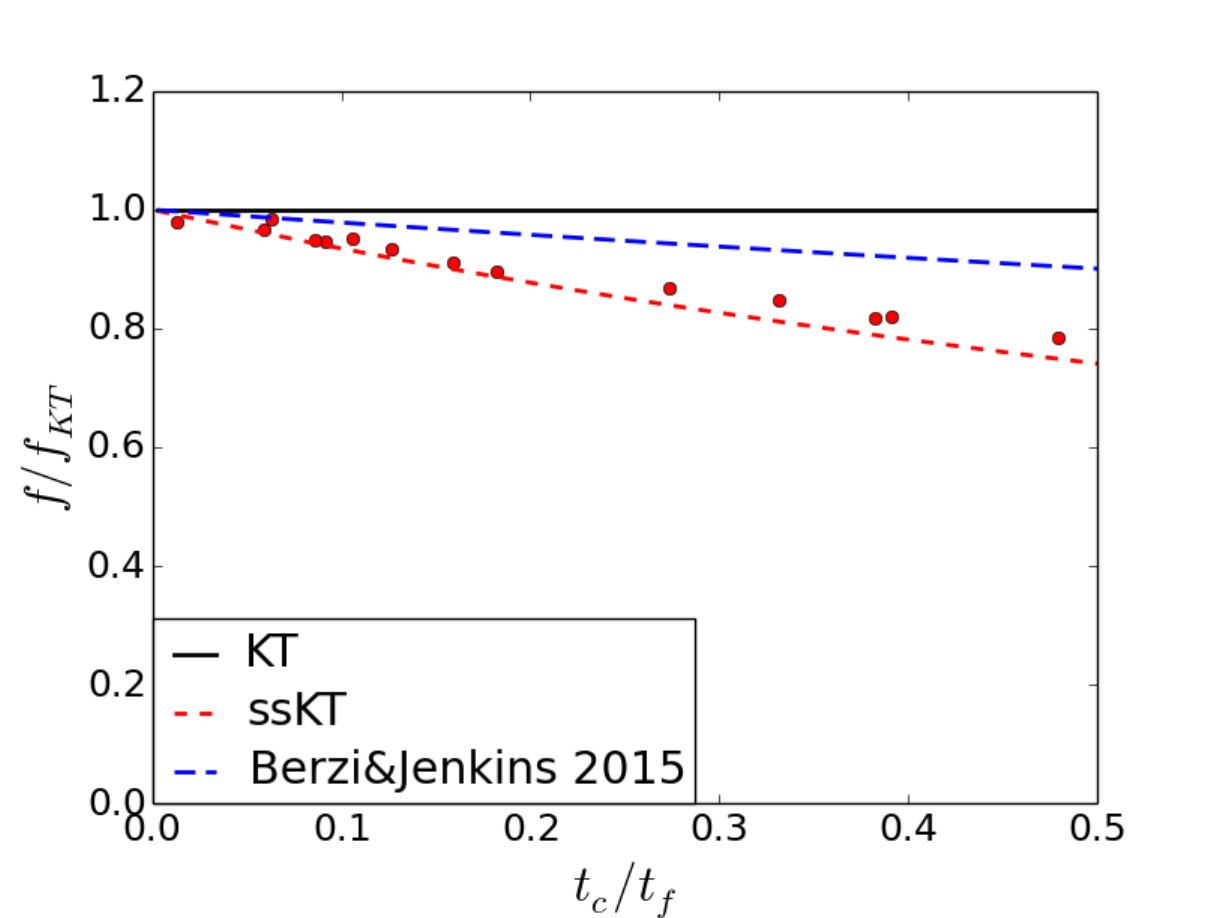}
\caption{Collision frequency $f$ (dots represent DEM results).The predicted collision frequency and pressure by \textcite{berzi2015steady} are also plotted for comparison.}
\label{fig:freq_comp}
\end{subfigure}\hfill
\begin{subfigure}{1\columnwidth}
\includegraphics[width=\columnwidth,keepaspectratio]{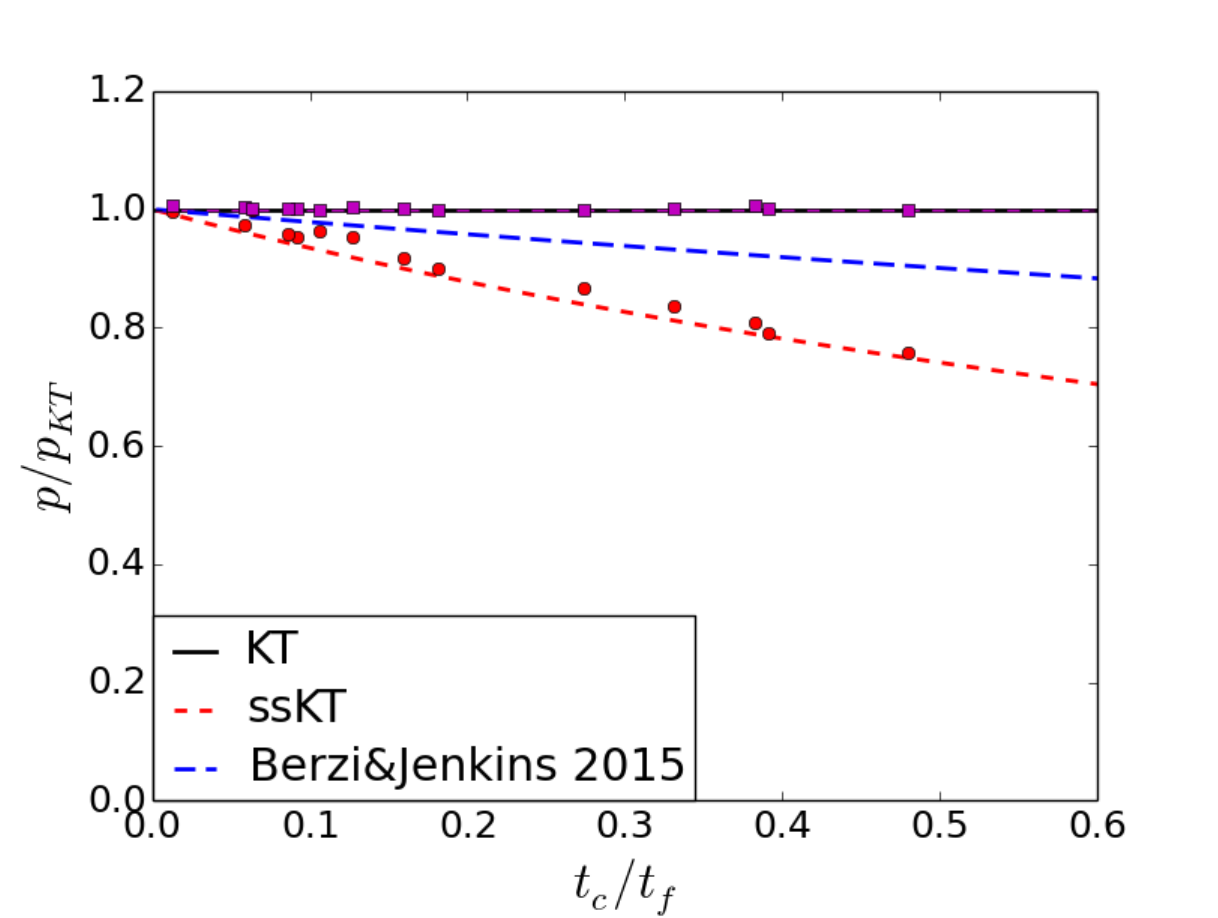}
\caption{$ p_{kin}$ and $p_{col}$ (dots represent $p_{col}$ and squares represent $p_{kin}$ from DEM results). The predicted collision frequency and pressure by \textcite{berzi2015steady} are also plotted for comparison.}
\label{fig:pres_comp}
\end{subfigure}
\caption{Comparison between the prediction of KT models and DEM simulation results.}
\end{figure}

To validate the proposed SSKT model, we compared the predicted collision frequency with the DEM simulation results. We performed several DEM simulations with different $t_c/{t_f}$ at $\phi=0.3$.  Our results were compared with the predictions from the model by \textcite{berzi2015steady} based on measured $T_k$ from DEM. 
As shown in figure \ref{fig:freq_comp}, the collision frequency predicted by the SSKT model could match the reduced collision frequency better as  $t_c/{t_f}$ increases. In general, the flow pressure could be divided into two parts, $p=p_{col}+p_{kin}$. Here, $p_{col}$ is related to the collision frequency and $p_{kin}$ is related to the particle kinetic fluctuation. The SSKT predicts that $p_{kin}$ is not affected by the elastic energy and $p_{col}$ varies along with the reduced collision frequency as shown in figure \ref{fig:pres_comp}. 

In the case of steady shear flows, the driving forces are converted to granular temperature through shear work, and ‘heat’ is dissipated through inelastic collisions. 
For dense steady-state shear flows, the velocity correlation as a result of the force chain has to be considered \citep{jenkins2006dense,jenkins2007dense}.

\begin{equation}
     \Gamma \frac{d}{L}=\tau \dot\gamma
     \label{eq:balance}    
\end{equation}

\begin{equation}
    L=\max \Big[\frac{2(1-e)}{15}(g_0-g_{0,f})+1, 1 \Big]
\end{equation}

Here $L$ is the correlation length which is a function of the solid volume fraction and restitution coefficient. $g_{0,f}$ is the value of radial distribution function at $\phi_f=0.49$. \citep{berzi2014extended}

Eq.(\ref{eq:balance}) can be further simplified and lead to $T_k \propto \dot \gamma^2$. We can show that for steady homogeneous shear flows $T_e$ remains constant, which does not enter the energy balance equation. The effect of reduced collision frequency $f/f_{KT}$ on both sides of Eq.(\ref{eq:balance}) is cancelled out as well. The particle stiffness would have no impact on the bulk behavior if the flows are at steady state and the energy diffusion is negligible. However, particle stiffness can affect the constitutive relation between stresses and kinetic granular temperature. If we let $p_{hard}$, $\tau_{hard}$, and $T_{hard}$ be the collisional pressure, shear stress, and kinetic temperature predicted by the classical KT model and let $p_{soft}$, $\tau_{soft}$, and $T_{soft}$ be the pressure, shear stress, and kinetic temperature predicted by the SSKT model, we have found ${T_{soft}}/{T_{hard}}=1$, ${p_{soft}}/{p_{hard}}=f/f_{KT}$, ${\tau_{soft}}/{\tau_{hard}}=f/f_{KT}$ and ${p_{soft}}/{\tau_{soft}}={p_{hard}}/{\tau_{hard}}$. Therefore, the change of particle stiffness of homogeneous shear flows will not change $T_k$ and $\tau/p$. This is evidenced in the DEM simulations by \textcite{vescovi2016merging} as shown in Table \ref{table2}. Comparing two simple shear DEM simulations at $\phi=0.6$ with $k/\rho d^3\gamma^2$ being equal to $10^3$ and $10^7$, respectively, the difference between the $\tau/p$  values is less than 2\% and the difference between the $T_k$ values than 11\%. However, the difference between the $p$ values is as large as 75\% and the difference between the $\tau$ values is 72\%. This shows that the particle stiffness has a significant impact on the pressure and shear stress fields of granular shear flows.

\begin{table}
   \caption{measurement of steady state quantities at $\phi=0.6$ from \textcite{vescovi2016merging}}
   \label{table2}	
   \begin{ruledtabular}
   \def\arraystretch{1.2}
   \begin{tabular}{p{1.3cm} p{1.3cm} p{1.3cm}p{1.3cm}p{1.3cm}}

$k/(\rho d^3 \dot\gamma^2)$ 
& $p/(\rho d^2 \dot\gamma^2)$
& $\tau/(\rho d^2 \dot\gamma^2)$
& $T_k/(d^2\dot\gamma^2)$
& $\tau/p$
\\
   \hline
$10^3$ 
& 11.14
& 3.26
& 0.58
&0.2926
\\
$10^7$ 
& 40.48
& 11.85
& 0.66
& 0.2927
\\

   \end{tabular}
   \end{ruledtabular}
\end{table}

\section{The effect of particle stiffness on granular flows}\label{sec:effect}

\subsection{Steady shear flows and regime maps}

\begin{figure*}
\centering
\begin{subfigure}{2\columnwidth}
\includegraphics[width=0.9\columnwidth,keepaspectratio]{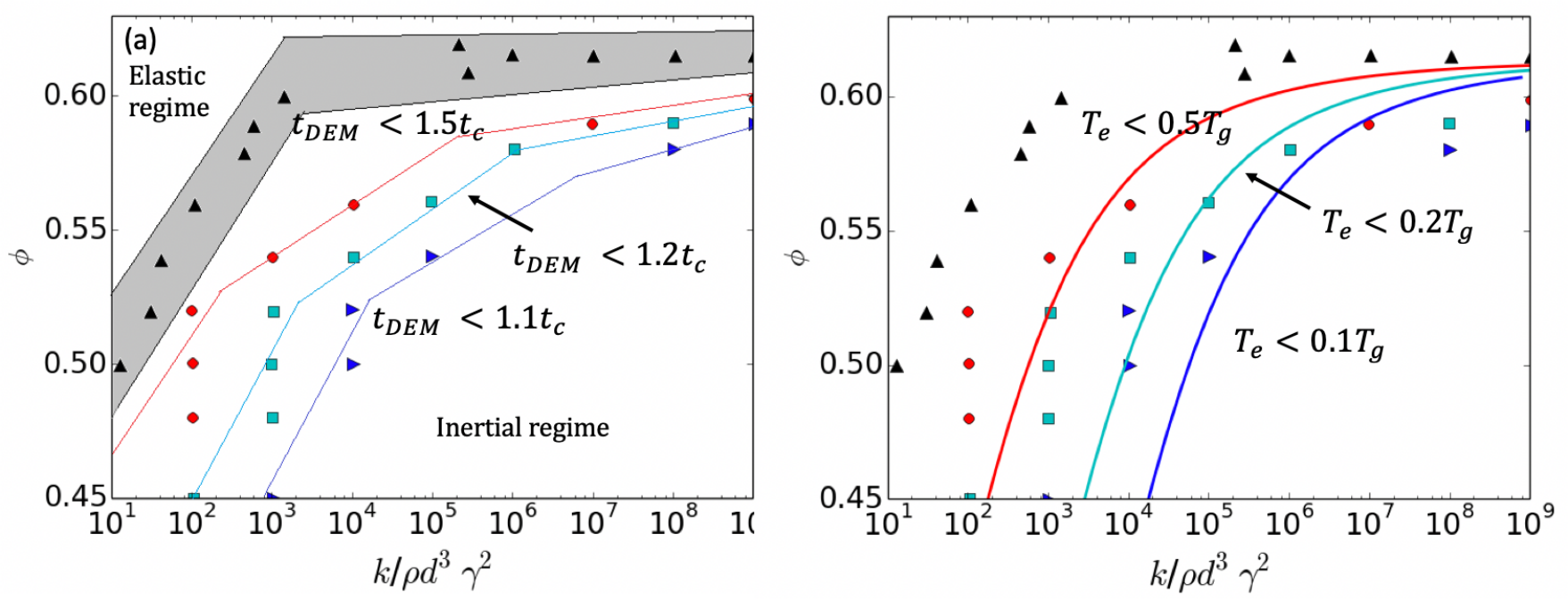}
\caption{$\mu=0.1$, left: fit with DEM data from Campbell; right: drawing based on SSKT}
\label{fig:6a}
\end{subfigure}\hfill
\begin{subfigure}{2\columnwidth}
\includegraphics[width=0.9\columnwidth,keepaspectratio]{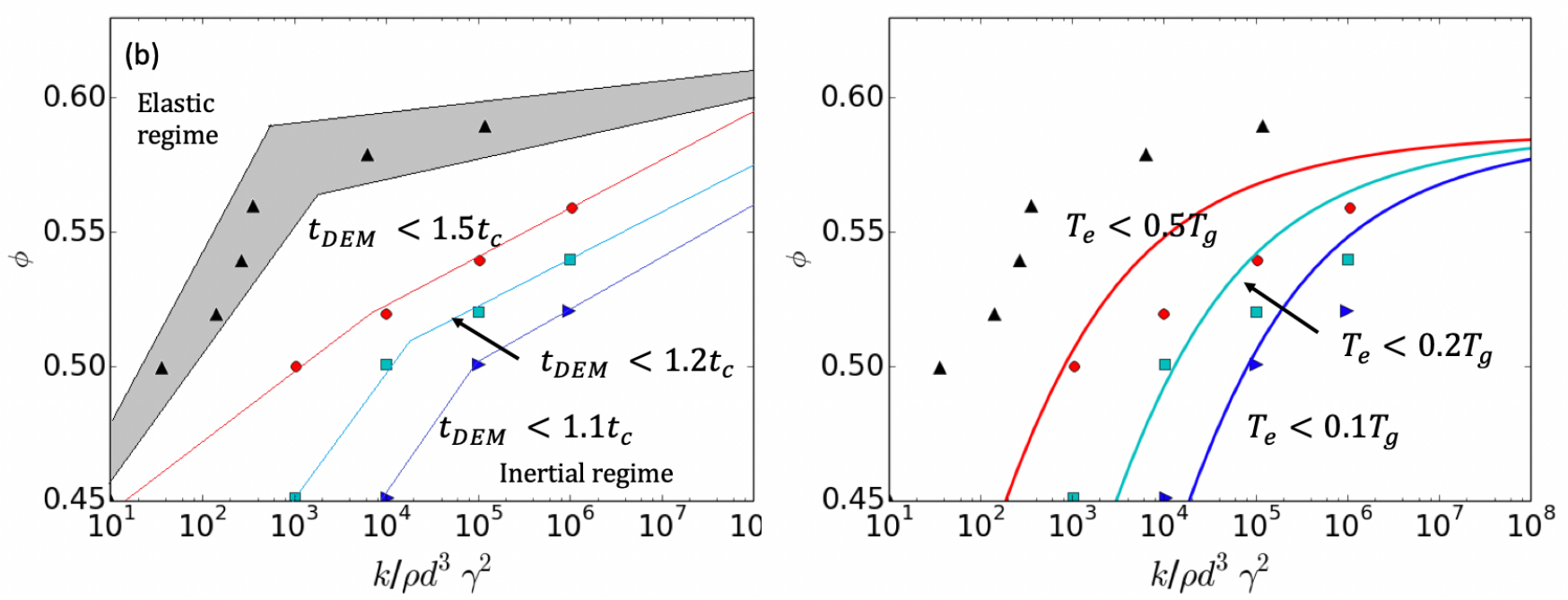}
\caption{$\mu=0.5$, left: fit with DEM data from Campbell; right: drawing based on SSKT}
\label{fig:6b}
\end{subfigure}
\begin{subfigure}{2\columnwidth}
\includegraphics[width=0.9\columnwidth,keepaspectratio]{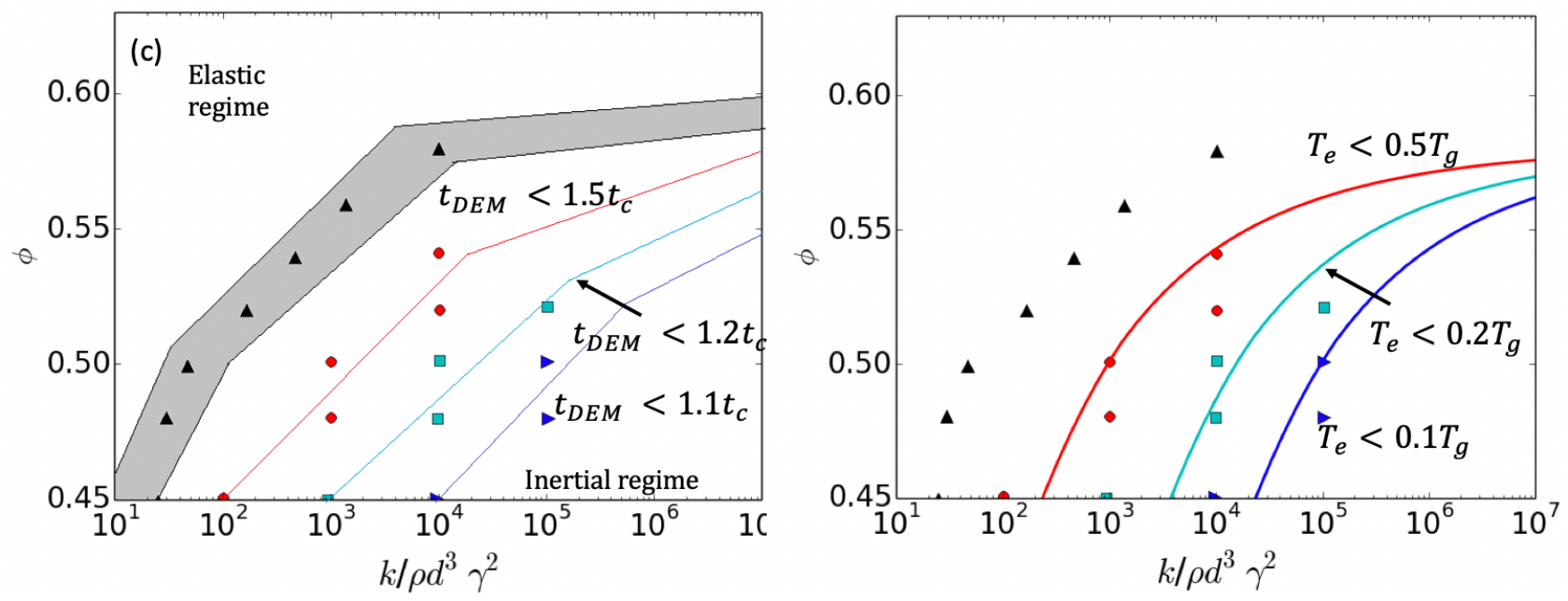}
\caption{$\mu=1.0$, left: fit with DEM data from Campbell; right: drawing based on SSKT}
\label{fig:6c}
\end{subfigure}
\caption{Regime maps as presented by Campbell (Campbell 2002) based on DEM simulations (left column) and presented model based on the ratio $T_e/T_k$ (right column). $T_e/T_k$=0.1: blue; $T_e/T_k$=0.2: green; $T_e/T_k$=0.5: red. }
\label{fig:6}
\end{figure*}

For steady homogeneous shear flows, $T_e$ remains constant, which does not enter the energy balance equation. Though $T_e$ of steady shear flows has no impact on the granular flow behavior, it can be used to indicate the granular flow regime and help quantitatively determine the boundaries on the regime map. 
Using data obtained from DEM simulations, many researchers constructed a regime map for granular flows to better understand the regime transition. Campbell unified the various granular flow theories and filled in the gap between the elastic granular flows and rapid granular flows. \cite{campbell2002granular} In his theory, the ratio of elastic effect to inertial effect is governed by a dimensionless parameter: $k/(\rho d^3 \gamma^2 )$. Since $k/(\rho d^3 \gamma^2 )\propto 1/(\gamma t_{c} )^2$, this ratio can be further interpreted as the square of the ratio of $1/\gamma$, a time scale that is relevant to how quickly particles are drawn together by the shear flow and how quickly the elastic contact forces push them apart\citep{campbell2006granular}. A complete flow map for shearing granular materials was presented based on the DEM simulation results as shown in the left column of figure \ref{fig:6}. His theory leads to the ironic conclusion that rapid granular flows (i.e., granular flows in the inertial regime) occur only at small shear rates when the solid volume fraction is fixed. He explained that increasing shear rate could compel the formation of force chains at lower concentrations. This phenomenon, on the other hand, could be explained by the SSKT model based on the role of elastic energy: the increasing shear rate could result in high temperature and high collision frequency, making the binary collision time comparable to the collision interval and invalidating the instantaneous collision assumption. The inertial-non-collisional regime on his map is equivalent to the transition regime as we discussed in this study; it indicates the limit where the KT starts to deteriorate as the elastic effect begins to dominate. The KT model becomes invalid as the flow moves further into the elastic regime. However, the transition regime on his regime map was estimated based on the DEM simulation results, and his theory did not provide an analytical method to accurately determine the extent of elastic effect in the transition regime. 

\textcite{sun2013energy} analyzed their DEM simulation results and found that the ratio of elastic energy to the kinetic energy could be used to determine the granular flow regime. Here we use the ratio $T_e/T_k$ from Eq.(\ref{3.18}) as the parameter to determine the significance of the elastic effect in the inertial regime. For frictional granular systems, we use the modification proposed by \textcite{chialvo2013modified}, which includes modified radial distribution function $g_{0,\mu}$ and energy balance equation between granular temperature and shear rate. 
\begin{equation}
	g_{0,\mu} = \frac{1-\frac{1}{2}\phi}{(1-\phi)^3} + \frac{0.58 \phi^2}{[\phi_c(\mu)-\phi]^{3/2}}
\end{equation}
where $\phi_c(\mu)$ is the $\mu$ dependent critical solid volume fraction \cite{chialvo2013modified}.

\begin{figure}
\centering
\begin{subfigure}{1\columnwidth}
\includegraphics[width=\columnwidth,keepaspectratio]{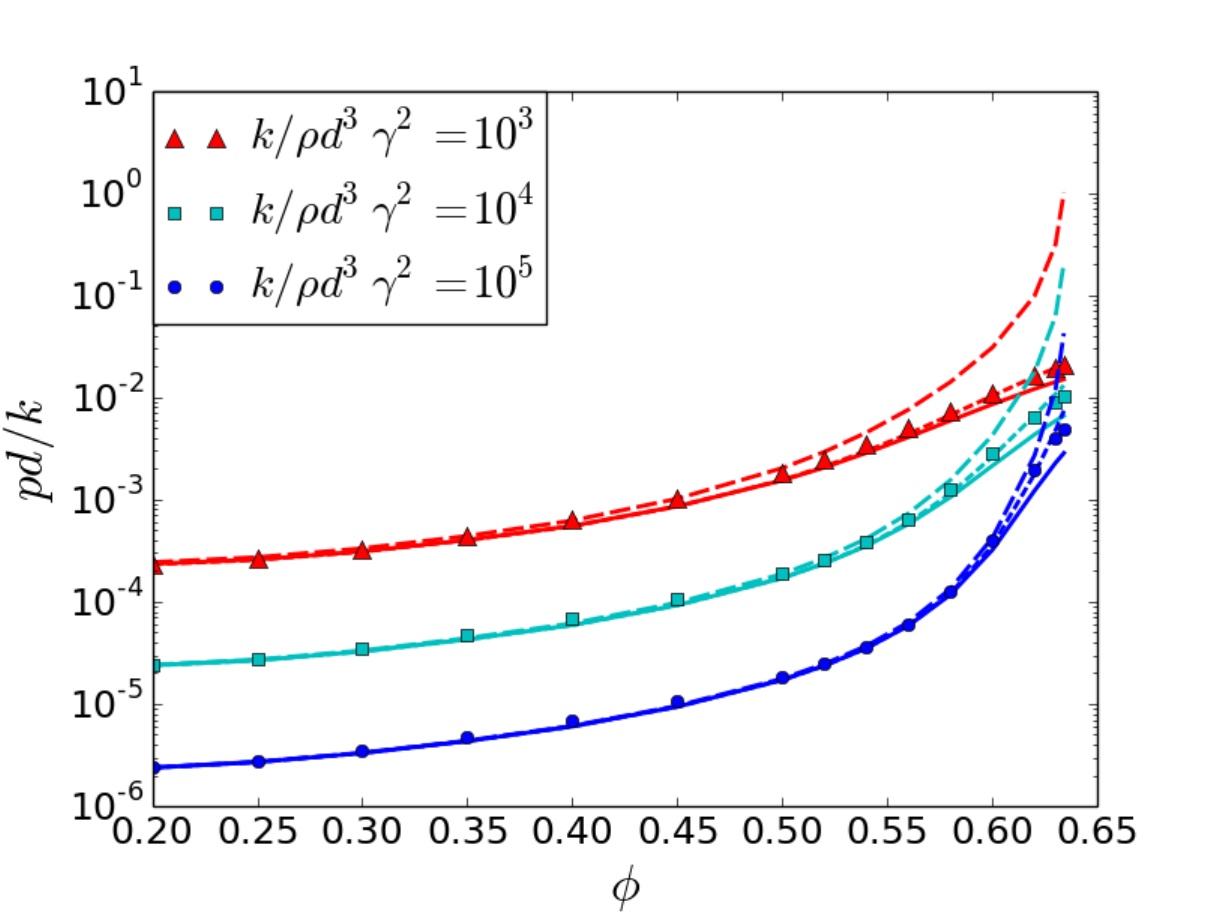}
\caption{scaled pressure vs. solid volume fraction $\phi$}
\label{fig:pressure}
\end{subfigure}\hfill
\begin{subfigure}{1\columnwidth}
\includegraphics[width=\columnwidth,keepaspectratio]{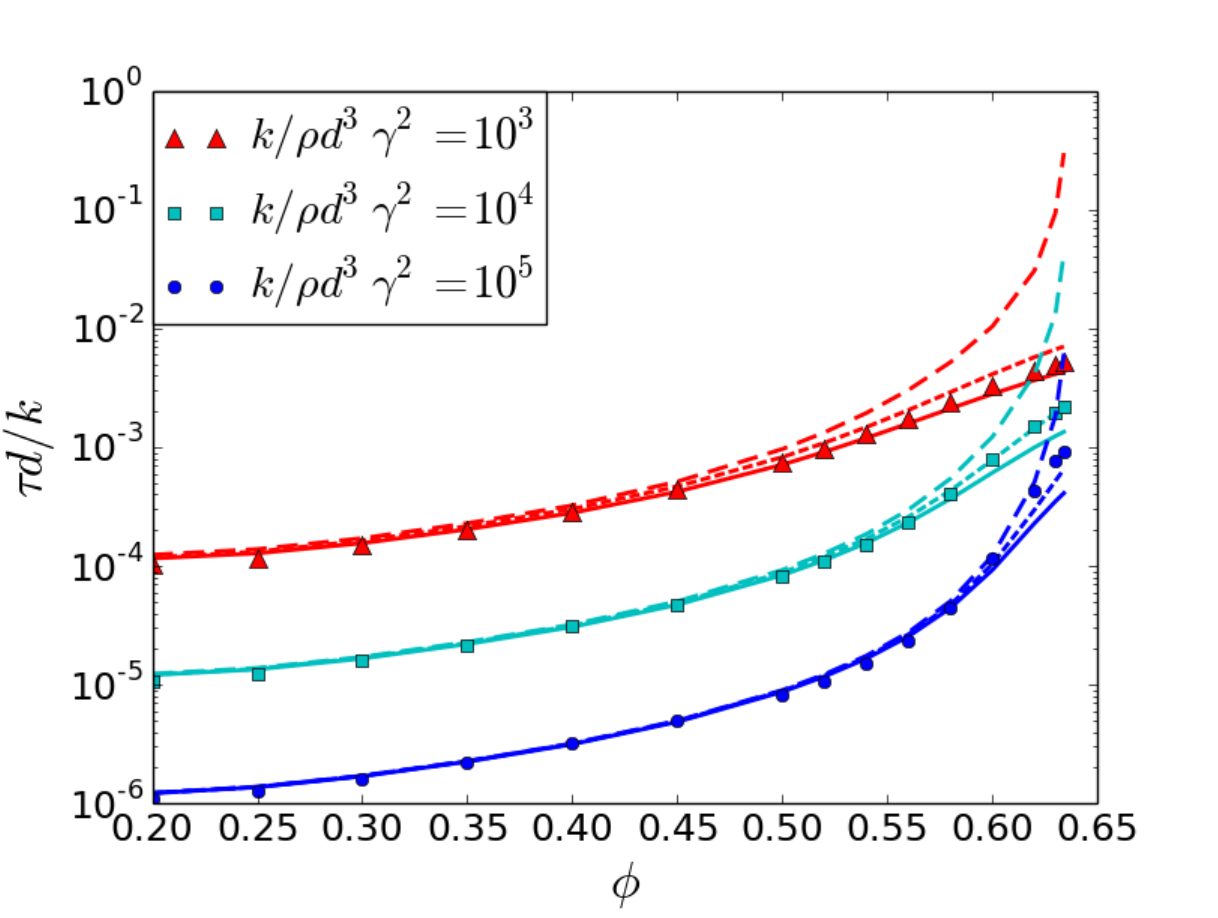}
\caption{scaled shear stress vs. solid volume fraction $\phi$}
\label{fig:shear}
\end{subfigure}
\caption{Comparison between theoretical predictions and results from DEM simple shear simulations. Dots represent DEM data from \textcite{vescovi2016merging}. Solid line represents the SSKT prediction. Short dashed lines and long dashed lines represent the model by \textcite{berzi2015steady} and the classical KT, respectively. }
\label{fig:vescovi}
\end{figure}

Based on the DEM simulation results, \textcite{campbell2002granular} calculated the ratio of averaged contact time from DEM to the theoretical binary collision duration determined by the particle properties, $t_{DEM}/t_{c}$, to determine if the particle interactions are collisional. If all the collisions are binary, the actual contact time $t_{DEM}$, must be equal to the binary collision duration, $t_{c}$. However, if the particles are involved in force chains in which particles endure longer contact time, one would expect the ratio $t_{DEM}/t_{c}>1$.  In theory, the KT model is only valid for collisional flows that corresponds to  $t_{DEM}/t_{c}=1$; any value larger than unity indicates non-binary collisional behaviour and the bulk behaviour would be affected by not only the granular temperature but also the elastic properties of particles. In his work, three lines were added to the regime map, roughly pointing to the places where $t_{DEM}/t_{c}$ falls below 1.5, 1.25, and 1.1, respectively. As the ratio becomes larger, the flow shows more elastic characteristics. A significant departure from the rapid flow behaviour is observed when $t_{DEM}/t_{c}=2$ at a solid fraction of 0.57 \citep{zhang1992interface}. Unlike their work in which the regime maps were drawn based on the DEM simulation results, here we quantitatively determine the extent of the elastic effect at the transition from inertial flows to elastic flows on the map based on the ratio $T_e/T_k$. Since $T_e/T_k$ is proportional to the fraction of multi-body collisions as discussed in the previous section, we expect the elastic granular temperature to be an alternative parameter that measures the elastic effect.  As shown in the right column of figure \ref{fig:6}, a very similar profile was observed as we change  $T_e/T_k$ from 0.1 to 0.5. This indicates that both $T_e/T_k$ and $t_{DEM}/t_{c}$ can be used to measure the extent of elastic effect in the system.

We also quantitatively compared our SSKT predictions with the DEM simulation results reported by \textcite{vescovi2016merging}. In their work, simulations under uniform shear were conducted for frictionless spheres with $k/\rho d^3 \gamma^2$ ranging from $10^3$ to $10^7$. We choose the cases with $k/\rho d^3 \gamma^2=10^3-10^5$ so that large amount of elastic potential energy exists. The measured kinetic granular temperature is then used as an input to calculate the pressure and the shear stress. As shown in figure \ref{fig:vescovi}, for flows at small solid volume fraction, $t_c/t_f$ is negligible because collision interval $t_f$ is large and the results from both KT model and SSKT model could match the DEM results well. However, as $\phi$ becomes large, $t_f$ will decrease and eventually become comparable to $t_c$; in this case the classical KT model over-predicts both pressure and shear stress and the discrepancy between the KT predictions and the DEM results becomes significant. It is also noticed that smaller $k$ results in larger $t_c/t_f$ and further increases the discrepancy.
On the other hand, our SSKT model could predict the pressure and shear stress very well compared with the DEM simulation results. We conclude that the inclusion of particle stiffness into the KT model is very important for modeling granular flows away from the inertial limit; the SSKT model is able to capture the reduced pressure and shear stress after considering the particle stiffness. 
The results from \textcite{berzi2015steady} model are also plotted in figure \ref{fig:vescovi}.  For steady flows, the elastic granular temperature remains unchanged; both models modify solid stresses through the reduced collision frequency. Even though a new expression of soft-particle collision frequency is used in the SSKT model, there is not much difference observed in the predicted pressures and stresses between SSKT model and Berzi and Jenkins model. However, it should be pointed out that a fitting contact duration $t_c=\frac{d}{5}(\frac{\rho \pi d}{4k})^{\frac{1}{2}}$ is used in Berzi and Jenkins model in order to match the DEM simulation results. This is not needed in our SSKT model since an accurate $t_{c}=\pi(\frac{k_n}{m_{eff}}-\frac{\eta_n^2}{4m_{eff}^2})^{-\frac{1}{2}}$ for LSD collision scheme has been used, which is consistent with the $t_c$ in DEM simulations. Compared with Berzi and Jenkins model, the SSKT model can also predict the transfer between kinetic and elastic energy, which is very important for transient flows that have varying $T_e$. The comparison between two models for unsteady cooling cases will be discussed in the next section.

\subsection{Homogeneous cooling cases}

For transient flows such as homogeneous cooling cases, $T_e$ changes along with time, which means part of the elastic potential energy is being converted to the kinetic fluctuation energy during the cooling process. The change of $T_k$ is a result of both inelastic collisions and energy transfer transfer from $T_e$.
The collision frequency is related to both solid volume fraction and the kinetic granular temperature; it increases as solid volume fraction or kinetic granular temperature increases, which means that the KT model can fail not only in the dense regime, but also in the dilute regime at high granular temperature. The applicability of KT model on uniform shear flows has been widely studied, but few studies have focused on the homogeneous cooling cases. As there is no driving force in a free cooling system, the initial granular temperature in the system will decrease due to inelastic collisions. Here we set up a three-dimensional free cooling system of fine particles. The system is dilute with a solid volume fraction $\phi=0.2$ and particle restitution coefficient $e=0.99$. We choose $e=0.99$ so that the high granular temperature will dissipate slowly and the effect of high collision frequency on soft-sphere systems can be more significant. The external force fields are ignored and only the particle energy dissipation is considered. The DEM results are compared with the theoretical predictions of the SSKT model to investigate its accuracy in terms of predicting the energy dissipation rate.
The change of total granular temperature $T_g$ is equal to the energy dissipation rate of the system $\Gamma$.
From Eq.(\ref{energyequation}) we have
\begin{equation}
       \frac{d}{dt}\big[\frac{3}{2}\rho\phi(T_k+\frac{1}{3}T_e)\big]=-\Gamma
       \label{cooling1}
\end{equation}
and from the SSKT model, we have
\begin{equation}
  	T_e=KH T_k^{{3}/{2}}
	\label{cooling2}
\end{equation}
Combing equations (\ref{cooling1}) and (\ref{cooling2}), the equation to solve $T_k$ becomes
\begin{equation}
     \frac{3}{2}\frac{dT_k}{dt}= \frac{12}{d\sqrt{\pi}}\phi g_0(1-e^2)\frac{T_k^{3/2}}{(1+\frac{1}{2} KH T_k^{1/2})[1+(1-\phi)     K T_k^{1/2}]}
\end{equation}
where $g_0(\phi)=\frac{1-\frac{1}{2}\phi}{(1-\phi)^3}$ is the radial distribution function for cases with $\phi<0.49$ \citep{carnahan1969equation}.

\begin{figure}
\centering
\begin{subfigure}{1\columnwidth}
\includegraphics[width=\columnwidth,keepaspectratio]{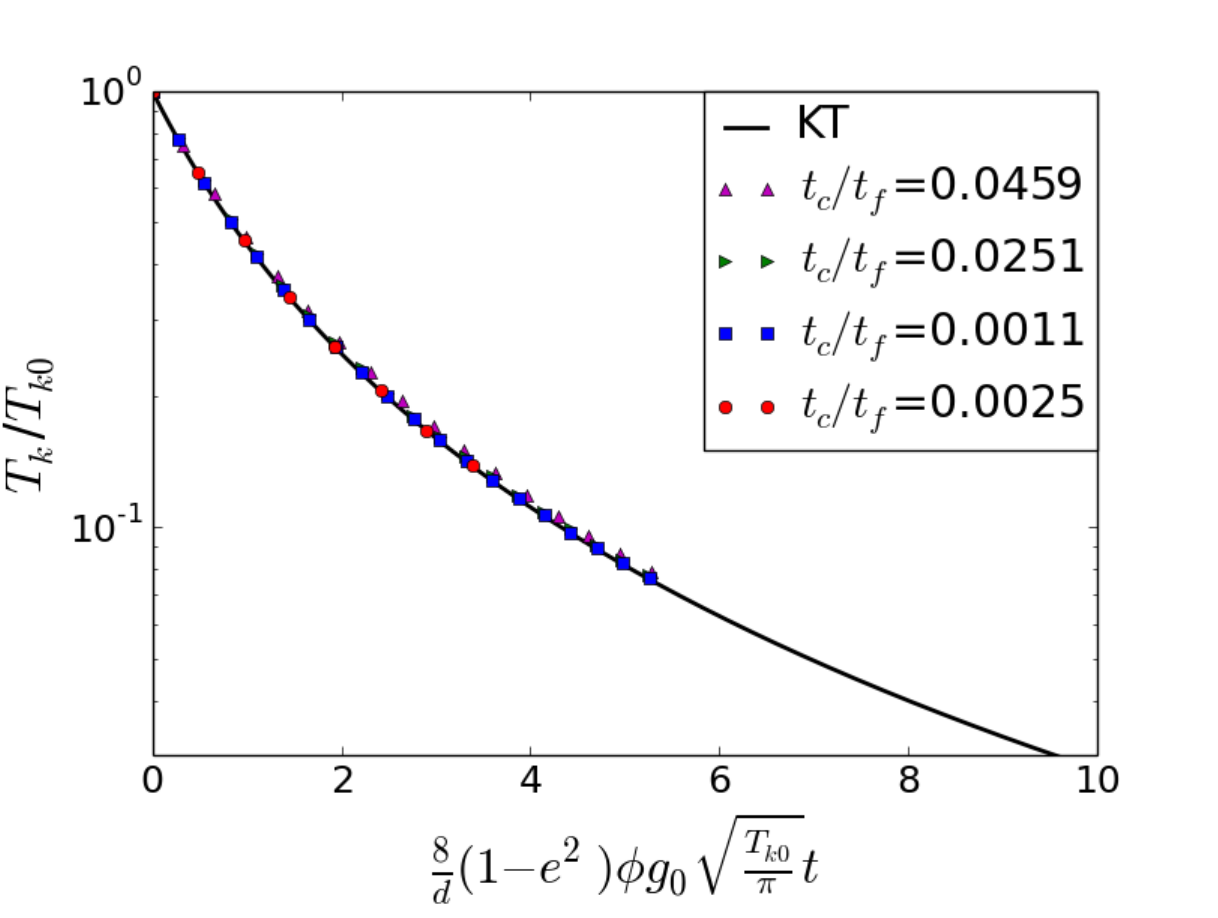}
\caption{$t_c/t_f<0.05$ at low initial granular temperature $T_{k0}=2\times 10^{-4}m^2/s^2$, different {\it k} has almost no impact on the results.}
\label{fig:2a}
\end{subfigure}\hfill
\begin{subfigure}{1\columnwidth}
\includegraphics[width=\columnwidth,keepaspectratio]{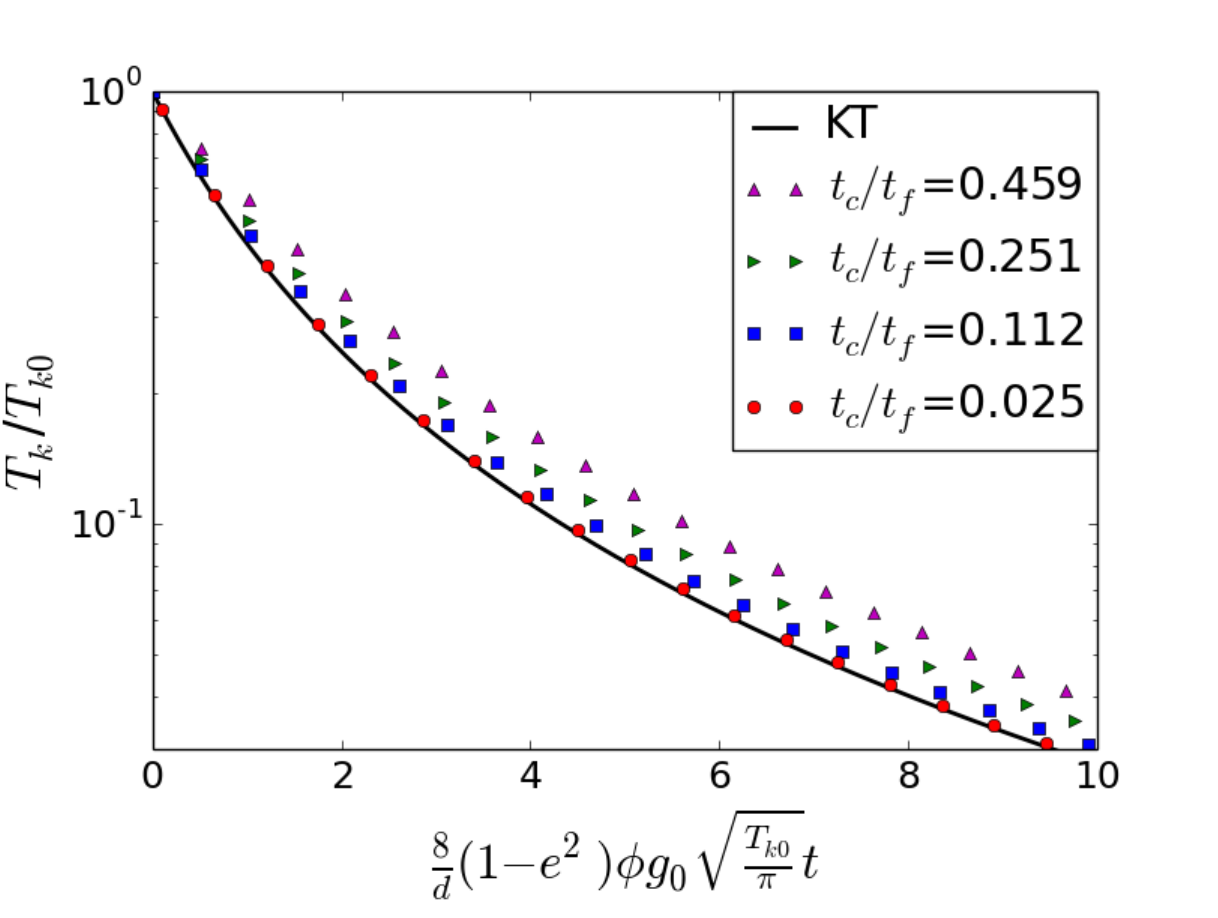}
\caption{$t_c/t_f>0.1$ at high initial granular temperature $T_{k0}=2\times 10^{-2}m^2/s^2$, different {\it k} has significant impact on the results if $T_{k0}$ is large.}
\label{fig:2b}
\end{subfigure}
\caption{ Kinetic granular temperature $T_k/T_{k0}$ as a function of rescaled time: $8/d (1-e^2 )\phi g_0 \sqrt{\frac{T_0}{\pi}} t$ when system is dilute and particles are nearly elastic($\phi=0.2,e=0.99$). The solid line shows the prediction of the classical KT and markers represent the DEM results.}
\end{figure}

\begin{figure}
\centering
\begin{subfigure}{\columnwidth}
\includegraphics[width=\columnwidth,keepaspectratio]{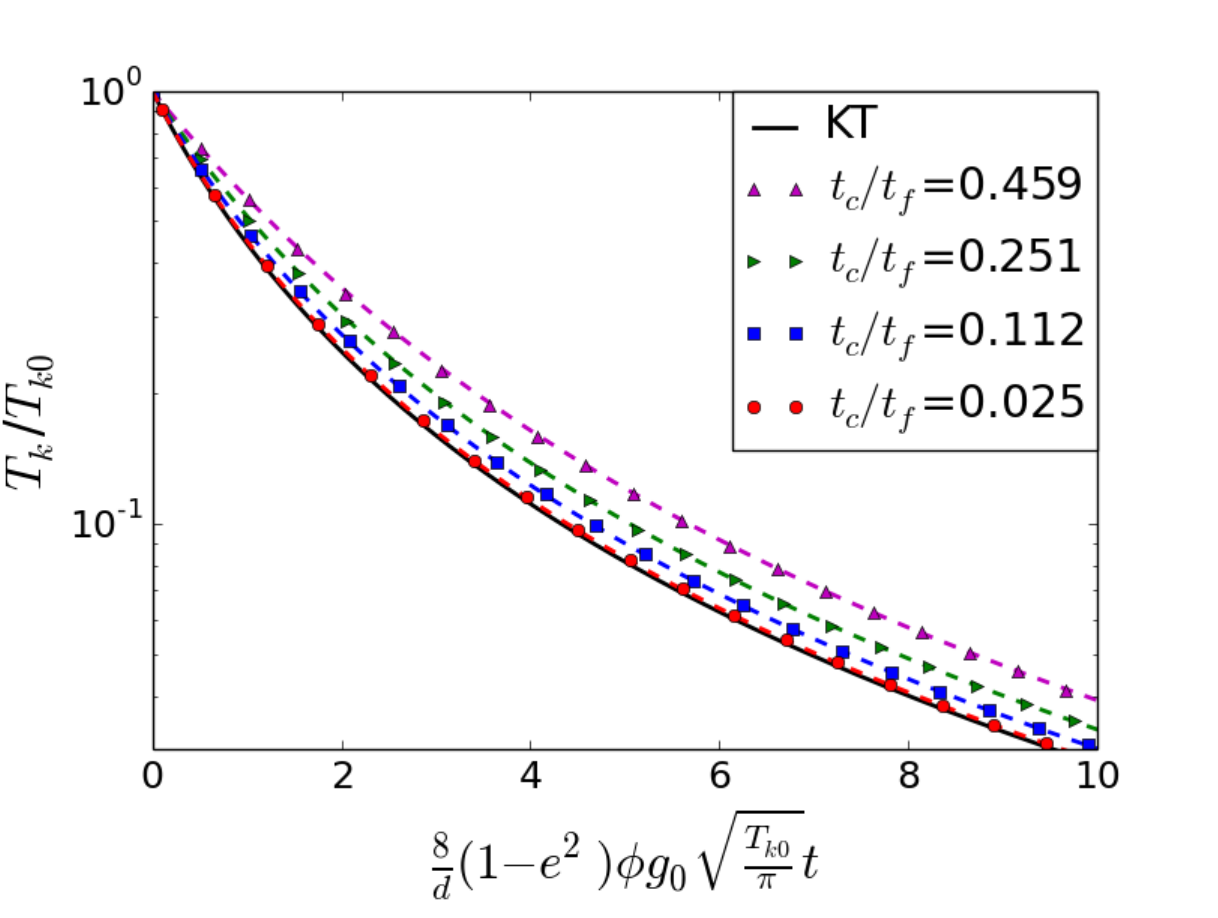}
\caption{ Kinetic granular temperature $T_k/T_{k0}$ as a function of rescaled time: $8/d (1-e^2 )\phi g_0 \sqrt{\frac{T_0}{\pi}} t$ when system is dilute and particles are nearly elastic. Solid line is the prediction of the classical KT and dashed lines are results of proposed SSKT model with different particle stiffness as input. Markers represent the DEM simulation data.}
\label{fig:2aa}
\end{subfigure}\hfill
\begin{subfigure}{\columnwidth}
\includegraphics[width=\columnwidth,keepaspectratio]{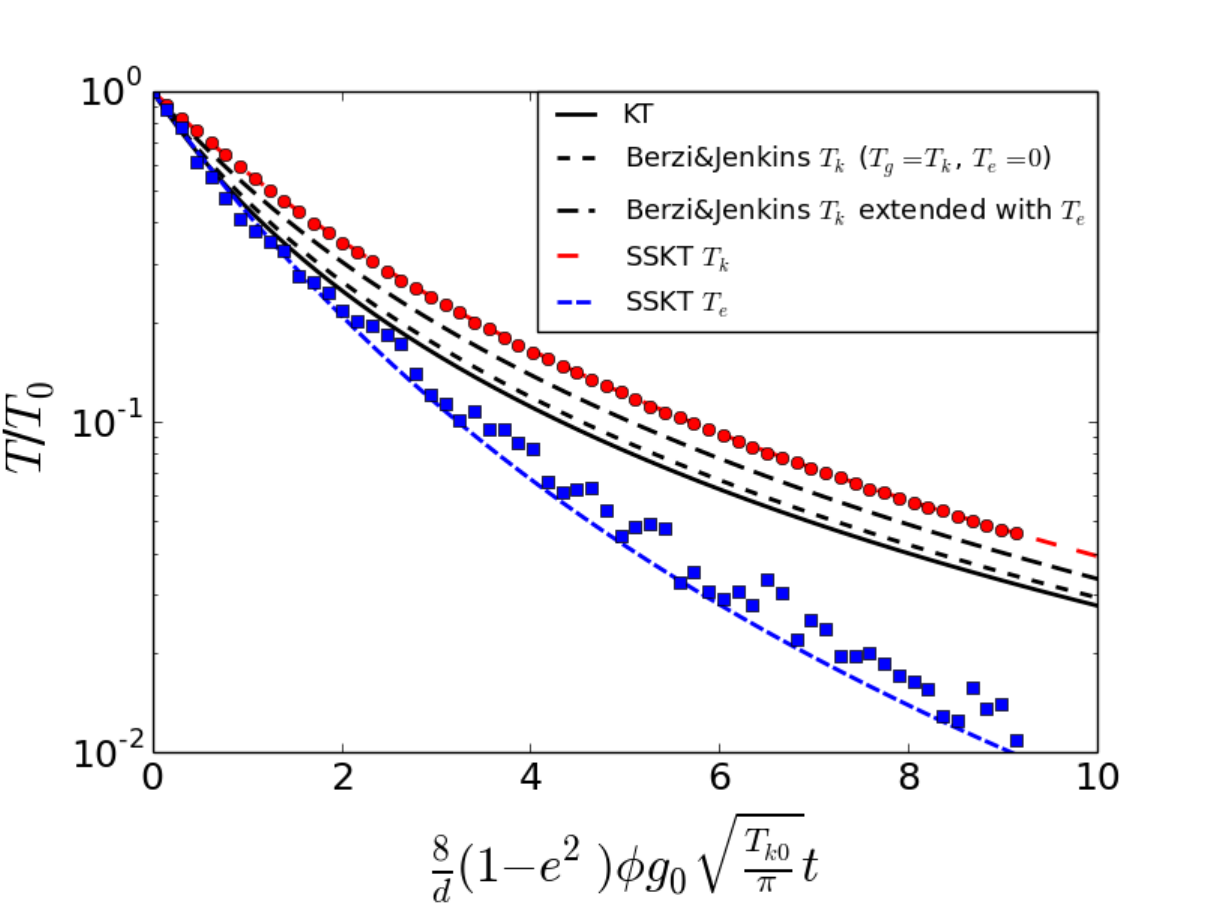}

\caption{ $T_k$ and $T_e$ predicted by SSKT compared with DEM results (circles: $T_k$; squares: $T_e$) and other models. $t_c/t_f=0.459$.}
\label{fig:2bb}
\end{subfigure}
\caption{Comparison between theoretical predictions and DEM results for homogeneous cooling cases.}
\label{fig:cooling}
\end{figure}

We first run the DEM simulation with low initial granular temperature $T_{k0}=2\times 10^{-4}m^2/s^2$ and vary particle stiffness $k$ from $10^2N/m$ to $10^4N/m$. This also changes the binary collision duration $t_{c}$. As shown in figure \ref{fig:2a}, the KT predictions match the DEM results well when the initial granular temperature is relatively low such that the initial $t_{c}/t_{f}$ is less than 0.05. During the cooling process, $t_{c}/t_{f}$ is further decreased due to the decay of $T_k$, making $t_c$ negligible. Contrary to the conventional wisdom, we find that the classical KT model can fail at low solid volume fractions if the particle collision frequency is sufficiently high (initial $t_c/t_f>0.1$), as shown in figure \ref{fig:2b}, in which the initial granular temperature $T_{k0}=2\times 10^{-2}m^2/s^2$. The KT prediction deteriorates because the increase of collision frequency causes the collision interval to decrease, making the collision duration a parameter that can no longer be neglected when compared to the collision interval. Therefore, a large $t_{c}/t_f$ at the initial stage increases the discrepancy between the predictions of the classical KT model and the DEM simulation results. As $t_{c}/t_f$ increases due to the decrease of particle stiffness $k$, particles become softer and a significant amount of kinetic energy is converted into non-dissipative elastic energy, resulting in a larger portion of elastic potential energy or a larger $T_e/T_k$ and a lower energy dissipation rate. At the same time, the collision frequency decreases as shown in Eq.(\ref{eq:f_soft}), and the energy dissipation rate is further reduced. From figure \ref{fig:2aa} we could see the classical KT can only match the result when $t_c/t_f$ is very small. On the other hand, the SSKT model is able to include the effect of particle stiffness and provide better predictions even at large $t_{c}/t_f$. 
In figure \ref{fig:2bb}, we also plot the results of the \textcite{berzi2015steady} model for the case of $t_c/t_f=0.46$. Realizing the velocity correlation should have no impact on flows without shear, we conclude that the energy dissipation rate of the homogeneous cooling flow is affected by the combination of reduced collision frequency and existence of non-dissipative elastic energy. Since the Berzi \& Jenkins model does not consider the effect of elastic potential energy, its prediction does not agree with the DEM simulations results. We extended Berzi \& Jenkins model by considering $T_g=T_k+T_e/3$ and using the collision frequency expression in their model. During the cooling process, large amount of elastic potential energy is converted to kinetic fluctuation energy. By including $T_e$ into the granular temperature $T_g$, it can significantly decrease the discrepancy between the DEM and the theoretical prediction. However, there is still difference between the predictions of the theory and the DEM because the reduced collision frequency is not accurately accounted for. On the other hand, by considering $T_g=T_k+T_e/3$ and using the proposed expression of collision frequency from Eq.(\ref{eq:f_soft}), the SSKT model is able to correctly predict the change of kinetic granular temperature as well as the elastic granular temperature during the cooling process.

\section{Conclusion} 

Finite particle stiffness makes sustained contacts possible. As a result, granular materials even in the dilute regime  can exhibit both collisional and non-collisional behaviors.
 Flow regimes have commonly been used to better understand the hydrodynamic behaviors of different kinds of granular flows. 
 For flows in the inertial regime, collisions can be treated as instantaneous events and the particle stiffness $k$ would have less impact on the results, so the classical KT model can accurately capture the flow behavior and compare well with the DEM simulations. However, as flows move away from the inertial regime to the transition regime, the particle stiffness becomes critical in predicting the characteristics of granular flows. In some cases, the collision interval predicted by the KT model can be comparable to or even smaller than the collision time, which can be interpreted as the existence of enduring contacts between particles. Therefore the modeling of granular flows in the transition regime has to consider the elastic properties of particles.  
In the present work, we have developed a SSKT model that takes the particle stiffness as an input parameter. 
The SSKT model extends the classical KT model by considering part of the kinetic fluctuation energy that actually exists in the form of elastic potential energy. This form of energy cannot be dissipated by collisions but can be transferred to the kinetic fluctuation energy spontaneously, which eventually affects the bulk behavior of granular flows. 
An elastic granular temperature is proposed to quantify the elastic potential energy in the system and a correlation between kinetic granular temperature and elastic granular temperature is derived based on the LSD collision scheme.
In addition, an expression for the reduced collision frequency is proposed based on the DEM simulation results, which could be used to modify the pressure and stress in the SSKT model and to explain the change of granular flow behavior in the transition regime. We have shown that the SSKT model is able to produce results that are in very good agreement with the DEM simulations. Considering the fact that granular flows of practical interest such as fluidized beds are unsteady and away from the inertial regime limit, the inclusion of particle stiffness into the continuum model allows us to improve the KT framework for a wider range of granular flows.

\section{acknowledgements}
This research work was supported by the U.S. Department of Energy.

\section{Appendix}

For three-dimensional systems the unperturbed particle distribution function could be written as
\begin{equation}
	f(\pmb c, \pmb r, t)=n\bigg( \frac{1}{2\pi T_k} \bigg)^{3/2} \exp\bigg(  -\frac{\pmb c^2}{2T_k} \bigg)
\end{equation}
where $n$ is the particle number density, $\pmb r$ and $\pmb c$ are the position and velocity of each individual particle, respectively.

The collisional integration for collision frequency is similar to the derivation of energy dissipation term, which can be found in the Appendix of another paper \cite{duan2017incorporation}.  Total number of collisions per second in the system takes the form

\begin{eqnarray*}
	\frac{dN_{col}}{dt} =\frac{d^2}{2} \int{d\pmb c_1} \int{d\pmb c_2}  \int_{\pmb c_{12} \cdot \pmb k <0}  (\pmb c_{12} \cdot \pmb k)g_0 n^2  \\ \bigg( \frac{1}{2\pi T_k} \bigg)^3
	 \exp\bigg(  -\frac{\pmb c_1^2 +\pmb c_2^2}{2T_k} \bigg)
 d \pmb k 
 \label{int}
\end{eqnarray*}

To obtain the integrations in Eq.\ref{int}, we need to transform the integral variables $d \pmb c_1 d\pmb c_2$ to $d \pmb c_{12} d\pmb c_{12}'$, where $\pmb c_1=\frac{\pmb c_{12}'+\pmb c_{12}}{2}$ and $\pmb c_2=\frac{\pmb c_{12}'-\pmb c_{12}}{2}$. The Jacobian of this transformation is $1/8$. 
We obtain 

\begin{equation}
	\frac{dN_{col}}{dt}=2\pi^{1/2}d^2 n^2 g_0 T_k^{1/2}
\end{equation}

The collision frequency is defined as the number of collisions per particle, so the we need to divide the number of collision by particle number density $n$, and $n=\frac{6\phi}{\pi d^3}$. 
Finally we have $f_{KT}$ 

\begin{equation}
	f_{KT}=\frac{1}{n} \frac{dN_{col}}{dt}=\frac{12}{d} \phi g_0 \sqrt{\frac{T_k}{\pi}}
\end{equation}

\nocite{*}
\bibliography{pof}

\end{document}